\documentclass[11pt]{article}

\usepackage[preprint]{acl}

\usepackage{times}
\usepackage{latexsym}
\usepackage{xspace}
\usepackage{amsthm}
\usepackage{booktabs} 
\usepackage{multirow} 
\usepackage{algorithm}
\usepackage{amsmath}
\usepackage{algorithmic}
\usepackage{changes}
\usepackage{listings} 
\usepackage{subcaption}
\usepackage{xcolor}
\usepackage{colortbl}

\usepackage{tcolorbox}
\usepackage{amssymb}
\usepackage{pifont}

\definecolor{astgreen}{RGB}{34, 139, 34}
\definecolor{astlightgreen}{RGB}{220, 255, 220}
\definecolor{textred}{RGB}{178, 34, 34}
\definecolor{textlightred}{RGB}{255, 230, 230}
\definecolor{codebg}{RGB}{248, 248, 248}
\definecolor{codeframe}{RGB}{200, 200, 200}

\tcbuselibrary{skins, breakable, listings}

\lstdefinestyle{bashstyle}{
    language=bash,
    basicstyle=\ttfamily\scriptsize,
    breaklines=true,
    backgroundcolor=\color{codebg},
    frame=single,
    rulecolor=\color{codeframe},
    xleftmargin=2pt,
    xrightmargin=2pt,
    aboveskip=4pt,
    belowskip=4pt,
}

\definecolor{poscol}{RGB}{34, 120, 34}   
\definecolor{negcol}{RGB}{192, 80, 77}   

\theoremstyle{definition}

\usepackage[T1]{fontenc}

\usepackage[utf8]{inputenc}

\usepackage{microtype}

\usepackage{inconsolata}

\usepackage{graphicx}

\usepackage{xcolor}
\newcommand{\codename}{\textsc{TrajEval}\xspace}

\title{Coherence Collapse: Diagnosing Why Code Agents Fail After Reaching the Right Code}

\author{
\textbf{Myeongsoo Kim$^{1}$, Dingmin Wang$^{1}$, Siwei Cui$^{1}$, Farima Farmahinifarahani$^{1}$,}\\
\textbf{Terry Yue Zhuo$^{2}$\thanks{Work done during an internship at AWS AI Labs.}, Shweta Garg$^{1}$, Baishakhi Ray$^{1}$,  Rajdeep Mukherjee$^{1}$, Varun Kumar$^{1}$} \\
$^{1}$AWS AI Labs \\
$^{2}$Monash University \\
\texttt{\{mysoo, wdimmy, siweicui, fafarima, shwegarg, rabaisha, mukherr, kuvrun\}@amazon.com} \\
\texttt{terry.zhuo@monash.edu}
}

\begin{document}
\maketitle
\begin{abstract}
Code agents resolve 65--70\% of SWE-bench Verified issues, but Pass@1 cannot tell us \emph{why} the rest fail---and, as we show, capable-model failures are systematically misdiagnosed without trajectory data. We introduce \codename, a training-free decomposition of agent trajectories into reference-patch-aligned search, read, and edit stages, and apply it across 16,758 trajectories spanning three architectures and seven models. The dominant failure of capable models is not localization: 60--69\% of failures on SWE-Agent and OpenHands reach \emph{and} edit the correct functions yet still produce incorrect patches, and the pattern persists for most models on the bash-only LiveSWEAgent. Within this Edit-Quality residual, we identify \textbf{Coherence Collapse}---the agent reaches correct code and then overwrites or thrashes it---as the largest theme, replicating across SWE-bench Verified and the multilingual PolyBench Verified. In 5 cases, the agent produces a patch \textbf{bit-identical to the gold reference} mid-trajectory and destroys it later; an edit-commit checkpoint recovers all 5 against the SWE-bench Docker harness. A reference-free consensus-driven variant yields a directional $+3.0$ pp Pass@1 measurement on GPT-5 ($p{=}0.08$).
\end{abstract}

\section{Introduction}
Code agents now resolve real GitHub issues at 65--70\% on SWE-bench Verified~\citep{jimenez2023swe,wang2024openhands,xia2025live}, yet the evaluation toolkit for them remains almost entirely outcome-based. Pass@1 treats each execution as a single binary outcome, mirroring the limitation that motivated process supervision for chain-of-thought reasoning~\citep{wei2022chain,huang2022language,schick2024toolformer}: aggregate correctness conceals \emph{where} and \emph{why} reasoning succeeds or fails. Code is well-suited to a process-level shift because a correct patch is a precise, automatically extractable specification of which files, functions, and code regions the model needed to identify and modify; comparing an agent's trajectory against this reference converts unstructured tool-use traces into stage-specific signals over localization, comprehension, and modification.

\begin{figure*}[!t]
\centering
\includegraphics[width=\textwidth]{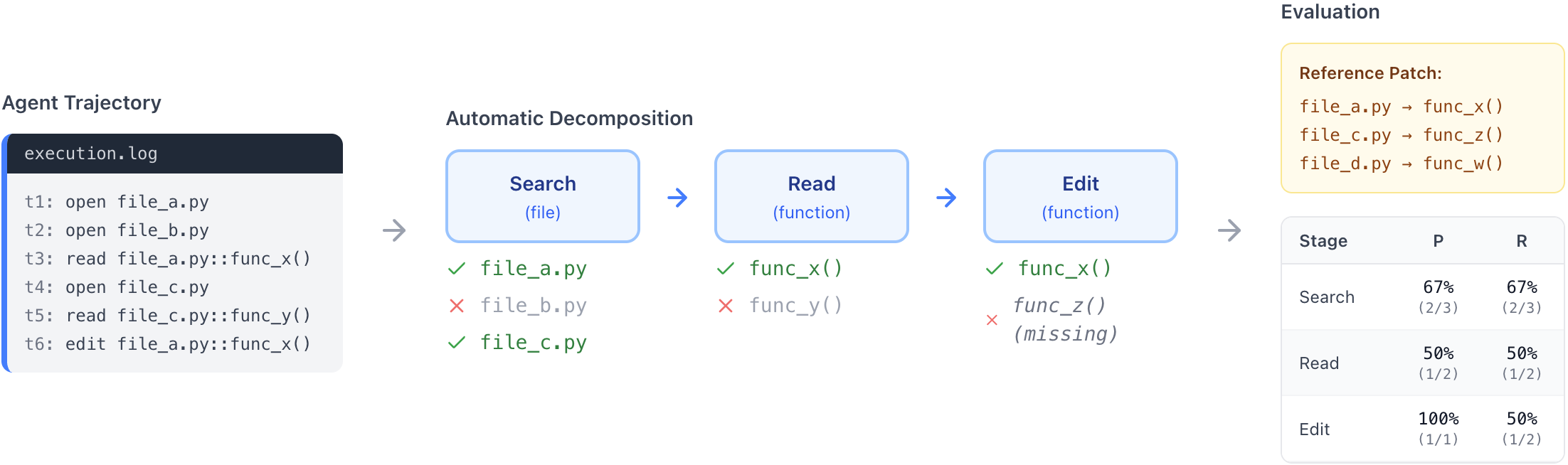}
\caption{Overview of \codename. Each agent execution trace is automatically decomposed into three stages---\textit{search} (file-level), \textit{read} (function-level), and \textit{edit} (function-level)---and compared against the reference patch to yield stage-wise precision and recall.}
\label{fig:overview}
\end{figure*}

We introduce \codename, a trajectory-based diagnostic framework that decomposes agent execution into three stages (Figure~\ref{fig:overview}): \emph{Search} (file-level), \emph{Read} (function-level inspection), and \emph{Edit} (function-level modification). Each stage yields precision (efficiency) and recall (effectiveness) against the reference patch.

Applied at scale to 16,758 trajectories across three agent architectures and seven models, this lens surfaces a finding outcome metrics cannot expose: 60--69\% of capable-model failures (GPT-5, GPT-5-mini, Qwen3-Coder-480B/30B, Qwen3-235B) on SWE-Agent and OpenHands reach \emph{and} edit the correct functions yet still produce incorrect patches, and the pattern persists for the GPT-5 family and Qwen3-235B on the bash-only LiveSWEAgent (Section~\ref{sec:failure_modes}). Within this Edit-Quality residual, the dominant pattern is \textbf{Coherence Collapse}: the agent reaches correct code and then overwrites or thrashes it, replicating across SWE-bench Verified and the multilingual PolyBench Verified, with a length-independent sub-type that dissociates from context-window degradation (Section~\ref{sec:edit_quality_taxonomy}, Appendix~\ref{sec:appendix_traj_length}). The finding narrows the explanation for capable-model failures: localization alone cannot account for them. Reference-patch-aligned trajectory analysis is emerging concurrently~\citep{ma2026p2t,dihan2026sweshepherd}; our contribution is to show that, at scale, it reveals an actionable failure mode---and that two interventions implied by the diagnosis (edit-commit checkpointing; parallel-sample consensus) admit direct end-to-end tests, reported in Section~\ref{sec:non_oracle}.

In summary, our contributions are:
\begin{itemize}
\item \textbf{A process-level lens for language-agent evaluation on code.} We introduce \codename, a training-free, architecture-agnostic procedure that decomposes an agent trajectory into search, read, and edit stages and computes precision and recall against a reference patch, operationalizing process supervision for code agents using a verifiable reference signal complementary to outcome-based Pass@k. We apply it at scale: 16,758 trajectories across three architectures (SWE-Agent, OpenHands, LiveSWEAgent) and seven models from 8B to 480B parameters on SWE-bench Verified and PolyBench Verified.

\item \textbf{Two findings with explicitly different scopes.} (i) \emph{Stage-level:} Edit-Quality is the modal failure mode for capable models on SWE-Agent and OpenHands (60--69\% of failures, all five capable models) and persists for the GPT-5 family and Qwen3-235B on the bash-only LiveSWEAgent (57--68\%) (Section~\ref{sec:failure_modes}). (ii) \emph{SWE-Agent only, sub-type taxonomy:} a 914-failure taxonomy identifies \emph{Coherence Collapse} as the largest theme on both SWE-bench Verified (39.7\%) and PolyBench Verified (32.3\%, including Java 36.3\% and JavaScript 42.1\%); inter-annotator agreement on the headline 3-class distinction is substantial ($\kappa = 0.80$, Appendix~\ref{sec:appendix_iaa}). Pass@1 progress for capable models is patch-quality-bound, not localization-bound.

\item \textbf{Two interventions of differing evidence quality.} The diagnosis predicts that edit-commit checkpointing should recover Near-Correct Corrupted failures and that parallel-sample consensus should approximate the oracle's confirmatory signal without reference-patch access. The first is a clean \emph{existence proof}: 5 intermediate edits bit-identical to gold pass the full SWE-bench Verified test suite when re-submitted to the Docker harness (Section~\ref{sec:non_oracle}). The second is a larger but \emph{directional} measurement, sub-threshold at $\alpha{=}0.05$: $+3.0$ pp Pass@1 on GPT-5 + SWE-Agent ($p{=}0.08$, $n{=}500$, Appendix~\ref{sec:appendix_consensus}).

\item \textbf{Reproducible artifact.} We release the trajectory-extraction pipeline, the per-trajectory feature CSV underlying every figure and table, and reproduction notebooks for each architecture~\citep{anonymous_trajeval_artifact_2026}.
\end{itemize}

\section{\codename: A Trajectory Analysis Lens}
\label{sec:approach}

We formalize agent trajectories as stage-specific sets over files, functions, and edit targets, then compare them against a reference patch to compute stage-wise precision and recall (Figure~\ref{fig:overview}). The procedure is lightweight, training-free, and architecture-agnostic, enabling consistent analysis across the three agent frameworks studied in Section~\ref{sec:experiments}.

\subsection{Problem Formulation}

Consider a code agent tasked with resolving an issue in repository $\mathcal{R}$ given an issue description $d$. The agent produces an execution trajectory
$T = \{(a_t, o_t)\}_{t=1}^{n}$, where $a_t$ denotes the action at step $t$ and $o_t$ the corresponding observation, and outputs a candidate patch $\hat{P}$.
Given a reference patch $P^*$, our goal is to extract interpretable features from $T$ that diagnose where agent behavior aligns with the reference solution.

The key insight is that $P^*$ provides a canonical reference for the modified context of the task: which files were changed, which functions were involved, and which code regions were modified.
We formalize this as the \emph{golden context}
$\mathcal{G} = (\mathcal{F}^*, \mathcal{H}^*)$,
where $\mathcal{F}^*$ denotes the set of files modified by $P^*$, and
$\mathcal{H}^*$ denotes the set of functions modified by $P^*$.

\subsection{Three-Stage Decomposition}

The three stages correspond to the three reference-patch projections the gold patch makes available---modified files, modified functions, and modified edit targets---yielding the coarsest decomposition that distinguishes navigation, comprehension, and modification without conflating them.
Let $\mathcal{F}_T$ denote the set of files viewed during trajectory $T$,
$\mathcal{H}_T$ denote the set of functions observed in agent observations, and
$\mathcal{H}_{\hat{P}}$ denote the set of functions modified in the candidate patch.

\paragraph{Stage 1: Search (File-Level).}
Search quantifies the agent's ability to localize relevant files within the repository.
Let $P_s$ and $R_s$ denote precision and recall for search:
\begin{equation}
P_s = \frac{|\mathcal{F}^* \cap \mathcal{F}_T|}{|\mathcal{F}_T|}, \quad
R_s = \frac{|\mathcal{F}^* \cap \mathcal{F}_T|}{|\mathcal{F}^*|}
\label{eq:search}
\end{equation}
High recall indicates successful file discovery, while high precision reflects focused exploration without redundant navigation.

\paragraph{Stage 2: Read (Function-Level).}
File localization alone is insufficient: an agent may open a file without examining the relevant code region.
Read quantifies function-level comprehension.
Let $P_r$ and $R_r$ denote precision and recall for read:
\begin{equation}
P_r = \frac{|\mathcal{H}^* \cap \mathcal{H}_T|}{|\mathcal{H}_T|}, \quad
R_r = \frac{|\mathcal{H}^* \cap \mathcal{H}_T|}{|\mathcal{H}^*|}
\label{eq:read}
\end{equation}
Read recall thus separates trajectories that examine the relevant function bodies from those that open a file but never inspect the buggy region.

\paragraph{Stage 3: Edit (Function-Level).}
Edit quantifies the localization accuracy of the agent's code modifications.
Let $P_e$ and $R_e$ denote precision and recall for edit:
\begin{equation}
P_e = \frac{|\mathcal{H}^* \cap \mathcal{H}_{\hat{P}}|}{|\mathcal{H}_{\hat{P}}|}, \quad
R_e = \frac{|\mathcal{H}^* \cap \mathcal{H}_{\hat{P}}|}{|\mathcal{H}^*|}
\label{eq:edit}
\end{equation}
Edit metrics capture \emph{localization} accuracy: an agent may achieve perfect edit recall while producing an incorrect patch.
Semantic correctness is evaluated separately via Pass@k.

\subsection{Golden Context}

The golden context averages 1.4 files and 2.1 functions per instance. We refer to this strict reference as \emph{Tier-0}; all precision and recall metrics in Sections~\ref{sec:experiments}--\ref{sec:results} are computed against it. We deliberately use a tight reference rather than a structurally expanded one (e.g., one-hop transitive dependencies via the call graph) so that low precision unambiguously reflects exploration beyond what the patch required, rather than artifacts of the reference's permissiveness. Appendix~\ref{sec:appendix_tier_sensitivity} reports a sensitivity analysis under a one-hop call-graph-expanded reference (\emph{Tier-1}) that supports this design choice empirically.

\paragraph{Reading the reference patch as a projection.} \codename{} measures agent behavior under a reference-patch projection: an action is counted as ``relevant'' if and only if it intersects code that the gold solution touches. We adopt this framing intentionally rather than treating the reference as ground truth on what an agent ``should'' do. Two observations support its informativeness as a projection. First, among 350 instances solved by two or more models---a selection-biased subset of easier instances, since unsolved instances offer no successful patches to compare---100\% show overlapping function-level modifications and 93.7\% show \emph{exact} function-level convergence (Section~\ref{sec:results}), indicating that on instances where comparison is possible, the modified-function set is largely invariant across successful agents rather than an arbitrary slice of the solution space. Second, the most actionable finding---Near-Correct Corrupted (Section~\ref{sec:edit_quality_taxonomy})---is validated by direct diff-hunk comparison rather than by the precision/recall thresholds, so its identification does not depend on the reference being uniquely correct. Alternative valid fixes that modify disjoint locations nonetheless register as low-recall, a strict-by-design tradeoff we discuss in Limitations~\ref{sec:limitations}.

\subsection{Feature Extraction}

Given an agent-specific tool mapping $\phi: \mathcal{A} \rightarrow \{\texttt{view}, \texttt{edit}, \texttt{other}\}$, extraction iterates over the trajectory and accumulates $\mathcal{F}_T$, $\mathcal{H}_T$, $\mathcal{H}_{\hat{P}}$ before computing Eqs.~\ref{eq:search}--\ref{eq:edit} in $O(|T|)$ time. Function extraction is hybrid (tree-sitter parsing with regex fallback) and supports Python, Java, JavaScript, and TypeScript. The full procedure is given in Appendix~\ref{sec:appendix_algorithm} (Algorithm~\ref{alg:extraction}).

\paragraph{What \codename{} does not measure.} \codename{} captures \emph{where} an agent operates, not whether the resulting code is correct. Two agents editing the same function with different patches receive identical edit metrics; debugging behaviors such as test interpretation, hypothesis formation, and post-edit verification appear only indirectly through their effect on stage-wise recall. The framework is designed to complement Pass@1, not replace it (Section~\ref{sec:limitations}).

\section{Experiments}
\label{sec:experiments}

\subsection{Experimental Setup}

\paragraph{Models.} We evaluate seven language models spanning 8B to 480B parameters: GPT-5, GPT-5-mini, Qwen3-8B, Qwen3-32B, Qwen3-235B, Qwen3-Coder-30B, and Qwen3-Coder-480B. The GPT-5 family is accessed via the OpenAI API; Qwen models via OpenRouter. All inference uses each provider's default decoding parameters; closed-API runs are not bit-deterministic, but our analysis pipeline uses fixed seed 42.

\paragraph{Benchmarks.} \textbf{SWE-bench Verified}~\citep{jimenez2023swe} contains 500 human-verified real-world GitHub issues from 12 Python repositories. \textbf{PolyBench Verified}~\citep{polybench} contains 382 multilingual instances across Python, Java, JavaScript, and TypeScript from 20 repositories. An instance is resolved (Pass@1$=1$) if the agent's submitted patch passes all \texttt{FAIL\_TO\_PASS} tests without regressing any \texttt{PASS\_TO\_PASS} test, evaluated via the official SWE-bench Docker harness.

\paragraph{Agent architectures.} We analyze three architectures that span the tool-primitive design spectrum, each run under its own default trajectory-budget and harness configuration: \textbf{SWE-Agent}~\citep{yang2024swe} (specialized \texttt{view}/\texttt{str\_replace}/bash), \textbf{OpenHands}~\citep{wang2024openhands} (\texttt{fsRead}/\texttt{fsWrite}/\texttt{executeBash}), and \textbf{LiveSWEAgent}~\citep{xia2025live} (bash-only). This diversity lets us isolate model-intrinsic behavioral patterns from scaffold artifacts.

\paragraph{Trajectory counts.} After excluding Qwen3-8B and Qwen3-32B from LiveSWEAgent due to persistent OpenRouter rate-limit/timeout errors with the bash-only harness (Appendix~\ref{sec:appendix_failure_modes}), the design yields \textbf{9,500} SWE-bench and \textbf{7,258} PolyBench trajectories, \textbf{16,758} in total. The central capable-model finding is unaffected: both excluded models are capability-bound and fail at upstream Search regardless of scaffold (Section~\ref{sec:failure_modes}).

Code and analysis pipelines are available at an anonymous repository link~\cite{anonymous_trajeval_artifact_2026}.

\subsection{Evaluation Protocol}

We conduct three diagnostic analyses, each described in detail at first use:
\begin{itemize}
    \item \textbf{Failure-mode partition} (Section~\ref{sec:failure_modes}): every failed trajectory is classified into one of four mutually exclusive stage-level buckets via cascading recall thresholds on $R_s, R_r, R_e$.
    \item \textbf{Edit-Quality taxonomy} (Section~\ref{sec:edit_quality_taxonomy}): the residual Edit-Quality bucket is decomposed via deletion-overlap, token-overlap, and edit-count signals; inter-annotator agreement is reported at three granularities (Appendix~\ref{sec:appendix_iaa}).
    \item \textbf{Falsifiable predictions} (Section~\ref{sec:non_oracle}): two interventions implied by the diagnosis are tested end-to-end---edit-commit checkpointing (5 EXACT intermediate edits re-submitted to the SWE-bench Docker harness) and parallel-sample consensus (GPT-5 re-executed under SWE-Agent on all 500 SWE-bench Verified instances with confirmatory feedback driven by $\textsc{Consensus}_3$ rather than the gold patch).
\end{itemize}

\section{Results}
\label{sec:results}

We present three behavioral findings. Section~\ref{sec:failure_modes} partitions every failed trajectory into mutually exclusive stage-level failure modes and shows that capable-model failures concentrate in the Edit-Quality bucket on SWE-Agent and OpenHands, with the pattern persisting for most capable models on the bash-only LiveSWEAgent. Section~\ref{sec:edit_quality_taxonomy} decomposes that residual into a five-theme taxonomy on the SWE-Agent scaffold and identifies \emph{Coherence Collapse} as the dominant pattern, replicated on PolyBench Verified. Section~\ref{sec:non_oracle} states two predictions implied by the diagnosis and reports calibrated evidence for each: an existence proof for edit-commit checkpointing and a directional, sub-threshold measurement for parallel-sample consensus. Cross-agent behavioral patterns are reported in Appendix~\ref{sec:appendix_cross_agent}.

\subsection{Failure Mode Analysis}
\label{sec:failure_modes}

To make the per-model behavioral signatures concrete, we partition every failed trajectory (Pass@1 = 0) into one of four mutually exclusive, exhaustive stage-level failure modes using a cascading 0.5 threshold on stage-wise recall: \textbf{Search} ($R_s < 0.5$, never opens a sufficient share of relevant files), \textbf{Read} ($R_s \geq 0.5$, $R_r < 0.5$, reaches relevant files but misses relevant functions), \textbf{Edit-Target} ($R_r \geq 0.5$, $R_e < 0.5$, reaches relevant code but modifies wrong locations), and \textbf{Edit-Quality} ($R_e \geq 0.5$, reaches \emph{and} edits the right locations yet produces an incorrect patch---a residual category no localization-stage signal can explain). Appendix~\ref{sec:appendix_threshold} re-buckets the same trajectories at $\tau \in \{0.3, 0.5, 0.7\}$ and shows that Edit-Quality remains the modal failure mode for all five capable models at every threshold (53--72\%); the qualitative conclusions below are stable to this choice. The partition uses the strict Tier-0 reference (functions actually modified by the gold patch); under a one-hop call-graph-expanded Tier-1 reference, the absolute bucket distribution shifts mass toward Search while the model rank ordering by recall is preserved ($\rho = 1.00$). Tier-0 is therefore the tier under which downstream-failure discrimination is well-defined, and we report it accordingly (Appendix~\ref{sec:appendix_tier_sensitivity}).

Table~\ref{tab:failure_sweagent_combined} reports the resulting distribution for SWE-Agent on SWE-bench Verified (top) and PolyBench Verified (bottom); appendix tables~\ref{tab:failure_openhands_swebench}--\ref{tab:failure_livesweagent_polybench} extend the same breakdown to the remaining four (scaffold, benchmark) cells.

\begin{table}[!htbp]
\centering
\small
\setlength{\tabcolsep}{2.3pt}
\caption{Failure-mode distribution among failed trajectories (Pass@1 = 0), SWE-Agent. Buckets are mutually exclusive and exhaustive; each row sums to 100\%. $n$ counts failed trajectories for which all six stage features could be extracted; trajectories where the agent terminated before producing parseable observations or edits are excluded, most strongly affecting weaker models.}
\label{tab:failure_sweagent_combined}
\begin{tabular}{lrcccc}
\toprule
\textbf{Model} & \textbf{$n$} & \textbf{Search} & \textbf{Read} & \textbf{E.-Tgt.} & \textbf{E.-Qual.} \\
\midrule
\multicolumn{6}{l}{\emph{SWE-bench Verified}} \\
GPT-5            & 163 &  6.7 &  4.3 & 20.9 & \textbf{68.1} \\
GPT-5-mini       & 162 &  8.6 &  8.0 & 14.2 & \textbf{69.1} \\
Qwen3-Coder-480B & 187 &  9.1 & 10.7 & 13.4 & \textbf{66.8} \\
Qwen3-Coder-30B  & 231 & 11.7 & 10.8 & 10.8 & \textbf{66.7} \\
Qwen3-235B       & 249 & 17.7 &  7.2 & 10.4 & \textbf{64.7} \\
Qwen3-32B        & 362 & 25.7 &  9.4 & 27.3 & \textbf{37.6} \\
Qwen3-8B         & 253 & \textbf{41.1} & 11.1 &  2.4 & 45.5 \\
\textbf{Weighted avg.} & 1607 & 19.3 & 9.0 & 14.8 & \textbf{56.9} \\
\midrule
\multicolumn{6}{l}{\emph{PolyBench Verified}} \\
GPT-5            & 208 & 18.8 & \textbf{41.3} &  8.7 & 31.2 \\
GPT-5-mini       & 216 & 27.8 & 30.1 &  8.8 & 33.3 \\
Qwen3-Coder-480B & 261 & 28.0 & \textbf{39.8} & 11.1 & 21.1 \\
Qwen3-Coder-30B  & 287 & 26.8 & \textbf{36.6} & 11.5 & 25.1 \\
Qwen3-235B       & 295 & \textbf{38.3} & 25.8 & 11.2 & 24.7 \\
Qwen3-32B        & 339 & \textbf{61.4} & 15.3 & 10.9 & 12.4 \\
Qwen3-8B         & 353 & \textbf{99.2} &  0.6 &  0.3 &  0.0 \\
\textbf{Weighted avg.} & 1959 & \textbf{47.0} & 25.0 & 8.7 & 19.3 \\
\bottomrule
\end{tabular}
\end{table}

\paragraph{Edit-Quality dominates capable models.} Across the five GPT-5, Qwen3-Coder, and Qwen3-235B variants, 64--69\% of failures fall in the Edit-Quality bucket: the agent reaches and edits the right functions, yet the generated patch is still incorrect. The pattern replicates across scaffolds: Edit-Quality is the modal failure mode for all five capable models on OpenHands (60.8--67.7\%, Table~\ref{tab:failure_openhands_swebench}) and for the GPT-5 family and Qwen3-235B on LiveSWEAgent (57.6--67.6\%, Table~\ref{tab:failure_livesweagent_swebench}); on the bash-only LiveSWEAgent harness, Qwen3-Coder-480B shifts to Read-dominant (40.3\%) and Qwen3-Coder-30B becomes Search-modal (33.4\%, narrowly above Edit-Quality at 33.2\%), suggesting tool-interface effects compound the underlying diagnosis on bash-only scaffolds.

\paragraph{Capability-bound models fail upstream.} Qwen3-8B's failures concentrate at Search (41.1\%), and PolyBench amplifies this pattern further (Table~\ref{tab:failure_sweagent_combined}, bottom): Search becomes the modal failure mode for every model except GPT-5, with the weighted Search rate rising from 19.3\% on SWE-bench to 47.0\% on PolyBench. Multilingual repositories appear to stress file discovery in particular, consistent with the per-tool reliability degradation reported in Section~\ref{sec:appendix_agents}.

\paragraph{Distinct per-model signatures.} Identical Pass@1=0 outcomes encode different failures across models (Edit-Target vs.\ Search), and tool-formatting friction compounds the picture: Qwen3-32B's 27.3\% Edit-Target rate co-occurs with degraded file-manipulation tool reliability, and Qwen3-235B's 26.5\% \texttt{submit} success rate is the most striking SWE-Agent toolkit anomaly (Appendix~\ref{sec:appendix_agents}).

\subsection{Qualitative Taxonomy of Edit-Quality Failures}
\label{sec:edit_quality_taxonomy}

The Edit-Quality bucket (56.9\% of all SWE-Agent + SWE-bench failures) is the dominant residual for every capable model but cannot be decomposed by stage-level features alone. We classify all 914 such failures into five themes comprising nine sub-types via a cascading-precedence rule on three trajectory signals (deletion overlap, addition-token overlap, gold-file edit count); the full rule and threshold-robustness check appear in Appendix~\ref{sec:appendix_taxonomy_rule}. We note upfront that Coherence Collapse and Semantic Error share a structural property---both reach the correct location and produce incorrect code---and are operationally (not mechanistically) distinguished by token overlap with the gold edit. The Near-Correct Corrupted sub-type is more crisply distinguished, being validated by exact diff-hunk reconstruction (Appendix~\ref{sec:appendix_near_correct_validation}); it is the most mechanistically grounded sub-claim in the taxonomy.

\begin{table}[!htbp]
\centering
\small
\setlength{\tabcolsep}{2pt}
\caption{Taxonomy of Edit-Quality failures (n=914, SWE-Agent on SWE-bench Verified). Sub-types are mutually exclusive (cascading precedence; Appendix~\ref{sec:appendix_taxonomy_rule}).}
\label{tab:edit_quality_taxonomy}
\begin{tabular}{llrr}
\toprule
\textbf{Theme} & \textbf{Sub-type} & \textbf{$n$} & \textbf{\%} \\
\midrule
\textbf{Coherence} & Confused Thrashing & 266 & 29.1 \\
\textbf{Collapse} & Near-Correct Corrupted & 97 & 10.6 \\
\addlinespace
\textbf{Semantic} & Wrong Branch Point & 170 & 18.6 \\
\textbf{Error} & Partial Understanding & 67 & 7.3 \\
 & Wrong API Call & 37 & 4.0 \\
 & Right Loc., Wrong Logic & 28 & 3.1 \\
\addlinespace
\textbf{Mis-loc.} & Wrong Function in File & 161 & 17.6 \\
\addlinespace
\textbf{Scope} & Incomplete Fix & 64 & 7.0 \\
 & Missing Edge Case & 8 & 0.9 \\
\addlinespace
\textbf{Exec.~Gap} & Understood But No Edit & 16 & 1.8 \\
\midrule
\textbf{Total} & & \textbf{914} & \textbf{100.0} \\
\bottomrule
\end{tabular}
\end{table}

\paragraph{Coherence Collapse (39.7\% of the 914 Edit-Quality failures, i.e., 22.6\% of all SWE-Agent failures).} The agent reaches the correct code, often editing it, but its final patch abandons or breaks its own earlier progress. Two sub-types: \emph{Confused Thrashing} (29.1\%)---$\geq 3$ edit attempts on the gold file, none persisting into the final diff; and \emph{Near-Correct Corrupted} (10.6\%)---an intermediate edit overlaps the gold patch ($> 50\%$ token overlap) and is subsequently overwritten or corrupted by a later edit. The two sub-types dissociate along trajectory length (Appendix~\ref{sec:appendix_traj_length}).

The Near-Correct Corrupted sub-type is the most actionable. An exact-hunk validation against the gold patch's diff structure (Table~\ref{tab:near_correct_validation}) shows that 30/97 (31\%) reconstruct the gold patch's hunks at the EXACT or NEAR level, and 96/97 (99\%) target the gold patch's code region rather than incidentally overlapping it. In the 5 EXACT cases, the agent produced a patch identical to the gold reference and then destroyed it later in the trajectory (Appendix~\ref{sec:example_exact_overwrite} walks through one such case in detail). These trajectories were not capability-bound: the model had already produced a correct fix, then deleted it.

\begin{table}[!htbp]
\centering
\small
\caption{Diff-hunk validation of the 97 Near-Correct Corrupted trajectories. The 5 EXACT cases ground the existence proof in Section~\ref{sec:non_oracle}; 30 strong-match (EXACT + NEAR) bound the recovery ceiling.}
\label{tab:near_correct_validation}
\begin{tabular}{lrr}
\toprule
\textbf{Classification} & \textbf{Count} & \textbf{\%} \\
\midrule
EXACT     &  5 &  5.2 \\
NEAR      & 25 & 25.8 \\
PARTIAL   & 66 & 68.0 \\
DIFFERENT &  1 &  1.0 \\
\midrule
\textbf{Total} & \textbf{97} & \textbf{100.0} \\
\bottomrule
\end{tabular}
\end{table}

\paragraph{Coherence Collapse is not long-context degradation.} The aggregate rate grows with length ($\rho = 0.32$, Q1 21.7\% $\rightarrow$ Q4 63.7\%), but sub-types dissociate sharply: Confused Thrashing drives this almost entirely ($\rho = 0.34$), while Near-Correct Corrupted is length-independent ($\rho = 0.001$, stable across quartiles). Longer trajectories produce more intervening steps before corruption but not greater probability of corruption (Appendix~\ref{sec:appendix_traj_length}).

\paragraph{Other themes and per-model patterns.} \emph{Semantic Error} (33.0\%, right location but wrong fix logic) is stable across models (27--37\%), suggesting task difficulty rather than capability. \emph{Intra-File Mislocalization} (17.6\%) edits the correct file but wrong function within it---a finer-grained failure than the inter-file Edit-Target bucket. \emph{Scope/Completeness} (7.9\%) peaks at 11--13\% in GPT-5/GPT-5-mini, and \emph{Execution Gap} (1.8\%) peaks at 4.4\% in Qwen3-32B. Per model (Appendix~\ref{sec:appendix_edit_quality_by_model}): capable models (GPT-5/mini, 31--33\% Coherence Collapse) maintain reasoning coherence at the cost of shallower fixes; Qwen3-32B (48.5\%) loses the thread mid-fix most often.

\paragraph{Reliability.} A second author independently labeled 100 SWE-bench Edit-Quality failures against the same cascading rule. Agreement on the headline 3-class distinction (Coherence Collapse / Semantic Error / Other) is substantial ($\kappa = 0.80$, 90\% raw agreement); refinement to the 5-class theme level remains substantial ($\kappa = 0.71$). Disagreements concentrate on the Scope/Completeness boundary (15/23 disagreements) and never affect the 5 Near-Correct Corrupted EXACT cases (Appendix~\ref{sec:appendix_iaa}).

\paragraph{Cross-benchmark replication on PolyBench.} To test whether Coherence Collapse is a SWE-bench- or Python-specific artifact, we replicate the taxonomy on the multilingual PolyBench Verified (SWE-Agent, n=455 Edit-Quality failures; Table~\ref{tab:cross_benchmark_main}). Coherence Collapse remains the largest single theme (32.3\%); JavaScript (42.1\%) exceeds the SWE-bench Python baseline (39.7\%) and Java (36.3\%) is comparable to it---narrowing the explanation: the pattern is not specific to Python or to the SWE-bench instance pool. The aggregate drop is driven by PolyBench Python (26.4\%, the lowest cell), implicating benchmark composition rather than multilingual difficulty (full taxonomy and per-language tables in Appendix~\ref{sec:appendix_polybench_taxonomy}).

\begin{table}[!htbp]
\centering
\small
\setlength{\tabcolsep}{3pt}
\caption{Cross-benchmark and per-language Coherence Collapse rate (\% of Edit-Quality failures), SWE-Agent. JavaScript on PolyBench (42.1\%) exceeds the SWE-bench Python baseline and Java (36.3\%) is comparable, indicating the pattern is not Python- or benchmark-specific. Full taxonomy in Appendix~\ref{sec:appendix_polybench_taxonomy}.}
\label{tab:cross_benchmark_main}
\begin{tabular}{lccccc}
\toprule
\textbf{Benchmark} & \textbf{Overall} & \textbf{Py} & \textbf{Java} & \textbf{JS} & \textbf{TS} \\
\midrule
SWE-bench (n=914) & 39.7 & 39.7 & --- & --- & --- \\
PolyBench (n=455) & 32.3 & 26.4 & 36.3 & \textbf{42.1} & 32.1 \\
\bottomrule
\end{tabular}
\end{table}

\subsection{Falsifiable Predictions from the Diagnosis}
\label{sec:non_oracle}

The diagnosis in Sections~\ref{sec:failure_modes}--\ref{sec:edit_quality_taxonomy} implies specific predictions about the effect of interventions, each detectable from the trajectory alone. The two Edit-Quality sub-buckets imply different intervention classes---Coherence Collapse is an agent-loop problem (edit-commit checkpointing, trajectory summarization); Semantic Error is a code-generation-quality problem (test-driven refinement, multi-sample selection)---and we test the agent-loop predictions end-to-end below.

\paragraph{Prediction 1: Edit-commit checkpointing recovers the Near-Correct Corrupted residual.} If an agent's intermediate edit passes the existing test suite, freezing it should prevent the corrupting edit that follows. Diff-hunk validation (Appendix~\ref{sec:appendix_near_correct_validation}) shows that 30 of 97 NCC cases (31\%) reconstruct gold-patch hunks at the EXACT or NEAR level, and 96/97 (99\%) target the gold patch's code region. We re-evaluated the 5 EXACT intermediate edits against the SWE-bench test harness: \emph{all 5 pass the full test suite}, establishing edit-commit checkpointing as a deployable existence proof rather than a hypothetical intervention. The realistic ceiling within NCC is bounded above by 30 (Appendix~\ref{sec:appendix_checkpoint_validation}); none of these signals require the reference patch at deployment.

\paragraph{Prediction 2: Parallel-sample consensus approximates the oracle's confirmatory signal.} We define the \emph{oracle} as a confirmatory feedback signal that, whenever the agent views a file present in the gold patch, injects a system message confirming that the file is gold-relevant; this fires on 490/500 instances and yields the $+4.6$ pp lift in Table~\ref{tab:consensus_endtoend}. The oracle requires the reference patch at execution time and is therefore non-deployable. The 93.7\% function-level convergence among successful agents (Appendix~\ref{sec:appendix_alt_solutions}) implies that when $N$ independent samples agree on an edit location, that location is, with high probability, gold-aligned. For each SWE-bench instance, we construct $\textsc{Consensus}_N$ from files viewed by $\geq N$ of the 6 non-GPT-5 models (leave-one-out, no reference patch). At $N{=}3$, $\textsc{Consensus}_N$ recovers 97.3\% (477/490) of the oracle's triggering cases.

We then measure consensus-driven feedback end-to-end: re-running GPT-5 on all 500 SWE-bench Verified instances under the same SWE-Agent harness with the confirmatory feedback trigger driven by $\textsc{Consensus}_3$ rather than the gold patch (Table~\ref{tab:consensus_endtoend}). The measurement is sub-threshold at $\alpha = 0.05$ ($+3.0$ pp Pass@1, 69.0\% vs.\ 66.0\%; McNemar's $p = 0.08$, n=500; ${\sim}50\%$ power for a $+3$ pp effect), but the effect direction matches the analytical $+4.5$ pp ceiling (under perfect-precision triggering) and the gain/loss decomposition (39 instances gained, 24 lost, net $+15$ instances) identifies a concrete mechanism for the residual gap.

That mechanism is the precision tradeoff: at $\textsc{Consensus}_3$ precision of 52.4\%, roughly half of triggered confirmations target non-gold files, occasionally misdirecting the agent---suggesting a higher-precision filter ($N{=}4$ or a learned re-ranker) as a direct improvement path. The chain-of-thought analogue is self-consistency decoding~\citep{wang2022self}, here on tool-use trajectories rather than token sequences; \citet{kim2026scaling} use parallel sampling differently (patch ranking, not file-level confirmatory feedback).

\begin{table}[!htbp]
\centering
\small
\setlength{\tabcolsep}{3pt}
\caption{Consensus measurement on GPT-5, SWE-Agent + SWE-bench (n=500). Oracle uses gold patch; $\textsc{Cons}_3$ uses $\geq 3$-of-6 non-GPT-5 model agreement (no reference-patch access).}
\label{tab:consensus_endtoend}
\begin{tabular}{lrr}
\toprule
\textbf{Condition} & \textbf{Pass@1} & \textbf{$\Delta$} \\
\midrule
Baseline       & 66.0\% (330) & --- \\
Oracle         & 70.6\% (353) & +4.6 ($p<.001$) \\
$\textsc{Cons}_3$ & 69.0\% (345) & +3.0 ($p=.08$) \\
\bottomrule
\end{tabular}
\end{table}

\section{Related Work}
\label{sec:related}

\paragraph{Positioning.} Reference-patch-grounded trajectory analysis is concurrent work, not novel to this paper. \citet{ma2026p2t} use the reference patch to score teacher trajectories on SWE-Gym for PRM training; \citet{dihan2026sweshepherd} train a lightweight PRM on SWE-bench. Our contribution is not the paradigm but its application as a training-free \emph{evaluation} procedure, and---more centrally---the diagnoses and validated interventions it yields: \emph{Coherence Collapse} as the dominant capable-model failure mode, 5 bit-identical-to-gold intermediate edits verified through the SWE-bench Docker harness, and a $+3.0$ pp end-to-end consensus measurement on GPT-5.

\paragraph{Process Supervision for Language Agents.}
Outcome-only metrics (Pass@k~\citep{chen2021evaluating}, reference-similarity~\citep{ren2020codebleu}) discard trajectory information, motivating process reward models~\citep{wei2022chain,huang2022language,schick2024toolformer}; \codename{} is the evaluation analogue, complementary to PRM-style supervision~\citep{ma2026p2t,dihan2026sweshepherd} rather than replacing it.

\paragraph{Code Agents and SWE-bench Resolution.}
SWE-Agent~\citep{yang2024swe}, OpenHands~\citep{wang2024openhands}, and LiveSWEAgent~\citep{xia2025live} span the design spectrum from specialized commands to structured primitives to shell-only, letting us isolate model-intrinsic behavioral patterns from tool-interface artifacts. Agentless~\citep{agentless} achieves competitive Pass@1 with a fixed three-phase pipeline (localization $\rightarrow$ repair $\rightarrow$ validation); our search/read/edit decomposition is reminiscent but applied \emph{descriptively} to any agent's trajectory rather than prescriptively as architecture.

\paragraph{Trajectory and Behavior Analysis for Agents.}
General agent benchmarks~\citep{liu2025agentbenchevaluatingllmsagents,zhou2024webarenarealisticwebenvironment,shi2017world} and code-domain studies~\citep{li2024devbench} focus on outcomes rather than within-trajectory structure. A recent wave of code-agent trajectory studies overlaps more directly with our premise. \citet{bouzenia2025thought} analyze thought-action-result patterns across RepairAgent/AutoCodeRover/OpenHands at the action-type level; \codename{} adds reference-patch-aligned precision/recall to attribute behavior to specific gold-context targets. \citet{majgaonkar2025understanding} examine SWE-bench failures across three scaffolds via outcome statistics; we partition failures into a stage-level taxonomy and, distinctively, validate the most actionable sub-type by re-executing 5 intermediate edits through the SWE-bench Docker harness---confirming each is bit-identical to the gold reference and passes the full test suite (Section~\ref{sec:edit_quality_taxonomy}, Appendix~\ref{sec:appendix_checkpoint_validation}). \citet{ma2026samesignal} run 64{,}380 SWE-bench executions testing cross-framework rule transfer; our scope is orthogonal---a per-trajectory diagnostic rather than a rule-transfer study---and we operate on three scaffolds with seven models against the reference patch. Concurrent work by \citet{sahoo2026agentlens} introduces ``Lucky Pass'' for trajectories that reach correct outcomes via incidental means; our \emph{Coherence Collapse} captures the dual phenomenon---correct intermediate state subsequently lost---with stage-level Edit-Quality dominance shown on SWE-Agent and OpenHands (and persisting for most capable models on LiveSWEAgent) and 5 EXACT cases re-executed against the SWE-bench Docker harness as an existence proof.

\section{Conclusion}
\label{sec:conclusion}

\codename{}---a reference-patch-aligned decomposition of agent trajectories into search, read, and edit stages---narrows the explanation for capable-model failures: localization alone cannot account for them. \emph{Edit-Quality} is the modal failure mode on SWE-Agent and OpenHands (60--69\%) and persists for the GPT-5 family and Qwen3-235B on LiveSWEAgent; within it, \emph{Coherence Collapse}---the agent reaches correct code and then overwrites or thrashes it---is the dominant SWE-Agent pattern (39.7\% on SWE-bench, 32.3\% on PolyBench), with a length-independent Near-Correct Corrupted sub-type that is not reducible to context-window degradation. Two interventions admit direct tests: an edit-commit checkpoint recovers 5 bit-identical-to-gold intermediate edits against the SWE-bench Docker harness as an existence proof, and parallel-sample consensus yields a directional $+3.0$ pp Pass@1 on GPT-5 ($p{=}0.08$) without reference-patch access. Immediate follow-ups: replicating the taxonomy on OpenHands and LiveSWEAgent, and validating trajectory summarization for the larger Confused Thrashing sub-type. We release the trajectory-extraction pipeline, per-trajectory features, and checkpoint-validation artifacts.

\section*{Limitations}
\label{sec:limitations}

\paragraph{Reference-patch dependence.} \codename{} requires a reference patch and is therefore not applicable at deployment time, and alternative valid fixes modifying different code locations would register as low-recall. The latter is partially mitigated empirically: among 350 multiply-solved instances, 100\% show overlapping function-level modifications and 93.7\% show exact convergence (Section~\ref{sec:results}). A reference-tier sensitivity analysis (Appendix~\ref{sec:appendix_tier_sensitivity}) shows that Tier-1 expansion ($11.7\times$ larger) collapses the failure-mode partition's discrimination while preserving model rank ordering ($\rho = 1.00$).

\paragraph{Decomposition is localization-only and not exhaustive.} The search/read/edit stages capture \emph{where} an agent operates, not \emph{whether the code is correct}: two agents editing the same function with different patches receive identical edit metrics, and behaviors central to debugging (bug reproduction, test interpretation, hypothesis formation, post-edit verification) appear only indirectly through their effect on stage-wise recall. \codename{} is intended to complement Pass@1, not replace it.

\paragraph{Consensus measured on one (model, scaffold) cell.} The end-to-end consensus measurement (+3.0 pp on GPT-5, $p = 0.08$) covers GPT-5 + SWE-Agent + SWE-bench Verified only; the same measurement on Qwen3-Coder-480B and on the OpenHands and LiveSWEAgent scaffolds is an immediate follow-up. The $p = 0.08$ result is below the analytical $+4.5$ pp ceiling (under perfect-precision triggering) and is sub-threshold at $\alpha = 0.05$; n=500 yields ~80\% power to detect a $+5$ pp effect and ~50\% power for $+3$ pp, so a larger benchmark or pooled measurement across scaffolds would tighten the confidence interval.

\paragraph{Edit-Quality taxonomy is heuristic-based, with multi-level IAA.} The Section~\ref{sec:edit_quality_taxonomy} taxonomy is computed from three trajectory signals (deletion overlap, addition-token overlap, gold-file edit count) under a cascading-precedence rule. Thresholds ($\delta=0.5$, $\tau=0.5$, $k=3$) were selected via single-annotator inspection of a 50-trajectory calibration set, fixed before a second author independently labeled 100 SWE-bench Edit-Quality failures (Appendix~\ref{sec:appendix_iaa}). Agreement on the headline 3-class distinction is substantial ($\kappa = 0.80$, 90\% raw agreement), also substantial at 5-class refinement ($\kappa = 0.71$), and lower-substantial on the full 6-category rule ($\kappa = 0.68$). Disagreements concentrate on Scope/Completeness (15/23) and never affect the 5 Near-Correct Corrupted EXACT cases, which are validated independently by diff-hunk reconstruction.

\paragraph{Cross-scaffold taxonomy replication is untested.} The Edit-Quality taxonomy is computed on SWE-Agent (SWE-bench and PolyBench) only; replication on OpenHands and LiveSWEAgent trajectories is left to future work.

\section*{Ethics Statement}
\label{sec:ethics}

This work analyzes execution trajectories from existing code agents on public benchmarks (SWE-bench Verified and PolyBench Verified) derived from open-source GitHub repositories under their original licenses. No human subjects, personal data, or private repositories were involved; all trajectories were produced by automated agents in sandboxed environments. The 16,758 trajectories were collected prior to this study; \codename{} itself runs on CPU in minutes.

We see two potential risks. First, the oracle signals in Section~\ref{sec:results} are non-deployable, and the end-to-end consensus measurement covers only a single (model, scaffold, benchmark) cell; we discourage interpreting either as guaranteed performance for end users. Second, stage-level diagnostics could in principle be used to game benchmark leaderboards by optimizing agents to match reference-patch trajectories rather than to solve underlying problems. We mitigate this by releasing the analysis pipeline as an evaluation tool rather than a training signal, and by explicitly distinguishing localization (where \codename{} operates) from semantic correctness (which Pass@1 measures and \codename{} does not replace).

\bibliography{custom}

\newpage
\appendix

\section{Appendix}
\label{sec:appendix}

\subsection{Appendix Overview}

This appendix provides supplementary results supporting the main paper findings:

\begin{itemize}
\item Cross-agent behavioral summary table and over-exploration metrics (Appendix~\ref{sec:appendix_cross_agent})
\item Cross-benchmark Edit-Quality taxonomy on PolyBench Verified, including per-language slice (Appendix~\ref{sec:appendix_polybench_taxonomy})
\item Robustness to alternative correct solutions: full convergence analysis (Appendix~\ref{sec:appendix_alt_solutions})
\item Complete efficiency metrics, token consumption, and tool success rates on SWE-bench and PolyBench
\item Trajectory feature extraction methodology
\item Edit-Quality taxonomy assignment rule and threshold robustness (Appendix~\ref{sec:appendix_taxonomy_rule})
\item Inter-annotator agreement on the cascading rule, reported at three levels of granularity (Appendix~\ref{sec:appendix_iaa})
\item Reference-tier sensitivity (Tier-0 vs.\ Tier-1; Appendix~\ref{sec:appendix_tier_sensitivity})
\item Failure-mode breakdowns for the remaining (scaffold, benchmark) cells
\item Near-Correct Corrupted diff-hunk validation (Appendix~\ref{sec:appendix_near_correct_validation})
\item Edit-commit checkpoint validation: 5/5 EXACT cases pass test suite (Appendix~\ref{sec:appendix_checkpoint_validation})
\item Consensus-as-relevance-signal validation (Appendix~\ref{sec:appendix_consensus})
\item Trajectory length and Coherence Collapse dissociation (Appendix~\ref{sec:appendix_traj_length})
\item Edit-Quality worked examples (Appendix~\ref{sec:appendix_worked_examples})
\end{itemize}

\subsection{Cross-Agent Behavioral Patterns: Full Tables}
\label{sec:appendix_cross_agent}

Table~\ref{tab:combined_behavior} reports the complete per-agent and per-model behavioral metrics referenced in Section~\ref{sec:results}.

\begin{table}[!htbp]
\centering
\small
\setlength{\tabcolsep}{4pt}
\begin{subtable}[t]{0.48\textwidth}
\centering
\begin{tabular}{lccc}
\toprule
\textbf{Metric} & \textbf{SWE-Agent} & \textbf{OpenHands} & \textbf{LiveSWE} \\
\midrule
Search Recall & 0.78 & 0.74 & 0.72 \\
Read Recall & 0.65 & 0.71 & 0.58 \\
Edit Recall & 0.68 & 0.52 & 0.45 \\
\midrule
Search Precision & 0.12 & 0.08 & 0.10 \\
Read Precision & 0.05 & 0.06 & 0.04 \\
Edit Precision & 0.25 & 0.18 & 0.15 \\
\midrule
Avg. Tool Calls & 30.7 & 24.2 & 13.5 \\
Files Viewed & 8.2 & 9.5 & 7.8 \\
\bottomrule
\end{tabular}
\caption{Cross-Agent Behavioral Summary.}
\label{tab:behavioral}
\end{subtable}
\hfill
\begin{subtable}[t]{0.48\textwidth}
\centering
\renewcommand{\arraystretch}{1.2}
\begin{tabular}{lcccc}
\toprule
\textbf{Model} & \textbf{Pass@1} & \textbf{Tools/Task} & \textbf{Submit \%} & \textbf{View \%} \\
\midrule
GPT-5 & 66.0\% & 30.7 & 99.9\% & 99.7\% \\
GPT-5-mini & 60.4\% & 32.2 & 99.9\% & 99.9\% \\
Qwen3-Coder-480B & 61.2\% & 50.3 & 99.9\% & 99.7\% \\
Qwen3-Coder-30B & 52.6\% & 50.5 & 100\% & 99.8\% \\
Qwen3-235B & 45.0\% & 42.5 & \textbf{26.5\%} & 99.9\% \\
Qwen3-8B & 13.2\% & 48.6 & 93.2\% & 99.2\% \\
Qwen3-32B & 14.8\% & 43.8 & 99.3\% & 100\% \\
\bottomrule
\end{tabular}
\caption{Efficiency Metrics.}
\label{tab:efficiency}
\end{subtable}

\caption{Behavioral patterns and efficiency metrics on SWE-bench Verified.}
\label{tab:combined_behavior}
\end{table}

Search recall averages 0.72--0.78 across agents, indicating that locating the correct files is a shared challenge regardless of tool implementation. Read recall shows more variance (0.58--0.71), with OpenHands achieving highest values, plausibly due to its structured file-reading interface. Edit recall ranges from 0.45--0.68, with SWE-Agent performing best, likely due to its specialized \texttt{str\_replace} command. Precision is uniformly low across stages and agents, indicating substantial over-exploration: agents inspect 8--12$\times$ more files and 17--25$\times$ more functions than necessary. The efficiency table also reveals deployment-relevant operational anomalies, most notably Qwen3-235B's 26.5\% submit success rate despite near-perfect view reliability; detailed tool-level analysis follows in Table~\ref{tab:app_swebench_success}.

\subsection{Edit-Quality Themes by Model}
\label{sec:appendix_edit_quality_by_model}

\begin{table}[!htbp]
\centering
\small
\setlength{\tabcolsep}{4pt}
\caption{Edit-Quality failure themes by model (\% of each model's Edit-Quality failures). Coh.~= Coherence Collapse; Sem.~= Semantic Error; Loc.~= Intra-File Mislocalization; Scope = Scope/Completeness; Gap = Execution Gap.}
\label{tab:edit_quality_by_model}
\begin{tabular}{lrrrrr}
\toprule
\textbf{Model} & \textbf{Coh.} & \textbf{Sem.} & \textbf{Loc.} & \textbf{Scope} & \textbf{Gap} \\
\midrule
GPT-5            & 33.3 & 34.2 & 20.7 & 10.8 & 0.9 \\
GPT-5-mini       & 31.2 & 32.1 & 22.3 & 13.4 & 0.9 \\
Qwen3-Coder-480B & 44.0 & 32.0 & 15.2 &  7.2 & 1.6 \\
Qwen3-Coder-30B  & 38.3 & 36.4 & 17.5 &  6.5 & 1.3 \\
Qwen3-235B       & 39.1 & 37.3 & 14.9 &  6.8 & 1.9 \\
Qwen3-32B        & \textbf{48.5} & 27.2 & 14.7 &  5.1 & \textbf{4.4} \\
Qwen3-8B         & 41.7 & 30.4 & 20.0 &  7.0 & 0.9 \\
\bottomrule
\end{tabular}
\end{table}

\subsection{Cross-Benchmark Taxonomy: PolyBench Verified}
\label{sec:appendix_polybench_taxonomy}

This section reports the full Edit-Quality taxonomy applied to PolyBench Verified, replicating the procedure used for SWE-bench Verified in Section~\ref{sec:edit_quality_taxonomy}. The same cascading-precedence rule (Appendix~\ref{sec:appendix_taxonomy_rule}) is applied to all 455 SWE-Agent + PolyBench Edit-Quality failures (the cell-specific count from Table~\ref{tab:failure_sweagent_polybench}'s Edit-Quality column).

\begin{table}[!htbp]
\centering
\small
\setlength{\tabcolsep}{4pt}
\caption{Edit-Quality theme distribution: SWE-bench Verified vs.\ PolyBench Verified (SWE-Agent scaffold). Coherence Collapse remains the largest single theme on both benchmarks; Semantic Error and Intra-File Mislocalization are within $\pm 1$ pp across benchmarks. The cross-benchmark drop in Coherence Collapse is driven by PolyBench Python (Table~\ref{tab:polybench_taxonomy_per_lang}) rather than by a multilingual effect.}
\label{tab:cross_benchmark_taxonomy}
\begin{tabular}{lrrr}
\toprule
\textbf{Theme} & \textbf{SWE-bench} & \textbf{PolyBench} & \textbf{$\Delta$} \\
 & (n=914) & (n=455) & (pp) \\
\midrule
Coherence Collapse         & 39.7\% & 32.3\% & $-7.4$ \\
Semantic Error             & 33.0\% & 33.9\% & $+0.9$ \\
Intra-File Mislocalization & 17.6\% & 16.9\% & $-0.7$ \\
Scope/Completeness         &  7.9\% &  8.3\% & $+0.4$ \\
Execution Gap              &  1.8\% &  8.6\% & $+6.8$ \\
\bottomrule
\end{tabular}
\end{table}

\paragraph{Per-language slice.} Table~\ref{tab:polybench_taxonomy_per_lang} reports the same theme distribution stratified by programming language within PolyBench Verified.

\begin{table}[!htbp]
\centering
\small
\setlength{\tabcolsep}{3pt}
\caption{PolyBench Edit-Quality taxonomy by language (SWE-Agent). JavaScript (42.1\%) exceeds and Java (36.3\%) is comparable to the SWE-bench Python baseline (39.7\%), contradicting the ``monolingual vs.\ multilingual'' hypothesis. The aggregate cross-benchmark drop is concentrated in PolyBench Python (26.4\%).}
\label{tab:polybench_taxonomy_per_lang}
\begin{tabular}{lrrrrrr}
\toprule
\textbf{Language} & \textbf{$n$} & \textbf{Coh.} & \textbf{Sem.} & \textbf{Loc.} & \textbf{Scope} & \textbf{Gap} \\
\midrule
Java       & 135 & 36.3 & 36.3 & 17.8 &  4.4 &  5.2 \\
JavaScript &  76 & \textbf{42.1} & 39.5 & 11.8 &  3.9 &  2.6 \\
Python     & 216 & 26.4 & 31.0 & 18.1 & 12.5 & \textbf{12.0} \\
TypeScript &  28 & 32.1 & 28.6 & 17.9 &  7.1 & \textbf{14.3} \\
\midrule
\textbf{All} & 455 & 32.3 & 33.9 & 16.9 & 8.3 & 8.6 \\
\bottomrule
\end{tabular}
\end{table}

\paragraph{Three observations.}

\emph{Coherence Collapse generalizes across languages and benchmarks.} JavaScript (42.1\%) exceeds and Java (36.3\%) is comparable to the 39.7\% SWE-bench Python baseline. Combined with the SWE-bench result, the phenomenon appears at $\geq 32\%$ in five independent (language, benchmark) cells, providing empirical evidence against the most natural alternative---that the finding reflects a Python or SWE-bench artifact.

\emph{The cross-benchmark drop is benchmark-composition, not multilingual.} The 7.4 pp aggregate drop in Coherence Collapse on PolyBench is concentrated in \emph{PolyBench Python} (26.4\%, 13.3 pp below SWE-bench Python). Java and JavaScript actually exhibit higher Coherence Collapse than the SWE-bench baseline. PolyBench Python issues differ in repository selection, issue type, and test structure from SWE-bench; these benchmark-composition factors---not language---drive the aggregate difference.

\emph{Execution Gap rises sharply on PolyBench, with two distinct mechanisms.} The Execution Gap theme (agent reads the gold file but never attempts an edit on it) jumps from 1.8\% to 8.6\%, concentrated in PolyBench Python (12.0\%) and TypeScript (14.3\%). We manually inspected 10 of the 16 PolyBench Python Execution Gap instances and identified two underlying mechanisms: (i) \emph{code paralysis} (4/10): the agent views the gold file 3--5 times, produces only reproduction scripts, and submits an empty or unrelated patch---concentrated on hard instances (\texttt{django-13513} recurs across 4 models); (ii) \emph{fix misattribution} (6/10): the agent reads the gold file thoroughly (mean 5.2 views) but concludes the fix belongs in an adjacent module---for example, on a \texttt{QuerySet.bulk\_create} bug in \texttt{query.py}, the agent edits \texttt{postgresql/operations.py} instead. Both mechanisms confirm the rise is not a parser artifact: Execution Gap is about reading the right code and failing to act on it, either through paralysis on hard instances or through misattribution to a related but incorrect module. This is consistent with PolyBench being curated to be challenging.

\paragraph{Scope.} Like the SWE-bench taxonomy in Section~\ref{sec:edit_quality_taxonomy}, the PolyBench taxonomy is computed only on the SWE-Agent scaffold; whether the cross-language pattern holds for OpenHands and LiveSWEAgent remains untested.

\subsection{Robustness to Alternative Correct Solutions}
\label{sec:appendix_alt_solutions}

To determine whether the reference patch represents the unique correct solution or whether alternative valid fixes may modify different code locations, we analyze 350 instances that were successfully resolved by two or more models. At the function level, 100\% of multiply-resolved instances show overlapping modifications---all models edit at least one common function. 93.7\% of instances show \emph{exact} function-level convergence, where all successful models modify precisely the same functions. At the file level, 99.4\% of instances show at least partial overlap, with only 0.6\% modifying completely disjoint file sets. This convergence suggests that SWE-bench issues have substantial convergence in modified locations, and---as we show in Section~\ref{sec:non_oracle}---this convergence is exploitable: a $\textsc{Consensus}_3$ relevance signal built from $\geq 3$-of-6 model agreement on viewed files yields a measured $+3.0$ pp Pass@1 lift on GPT-5 without reference-patch access (Appendix~\ref{sec:appendix_consensus}).

We observe 106 resolved instances (8.6\% of successful resolutions) with low edit recall ($<$0.8). Manual inspection of 10 cases with edit recall $=0$ reveals two patterns: (1) \emph{partial completions} where agents modify a subset of required functions (e.g., editing 2 of 4 golden locations), and (2) \emph{alternative solutions} where agents fix the bug via completely different code paths (e.g., editing \texttt{contenttypes/fields.py} when the reference patch modifies \texttt{fields/\_\_init\_\_.py}). Both patterns successfully resolve the issue by passing all tests.

\subsection{Complete Model Results: SWE-bench Verified}
\label{sec:appendix_agents}

\subsubsection{Efficiency Metrics}

Table~\ref{tab:app_swebench_efficiency} presents complete efficiency metrics for all seven models evaluated on SWE-bench Verified. GPT-5 family demonstrates a clear efficiency gradient: GPT-5 achieves the lowest average tool usage at 30.71 calls per task. Qwen3-235B achieves comparable efficiency to GPT-5-mini despite lower bash usage (15.25 vs 16.84), while Qwen3-Coder-30B and Qwen3-Coder-480B show elevated tool counts (50.45 and 50.30) with moderate localization percentages (20.5\% and 23.4\%), indicating more deliberate but verbose problem-solving strategies.

\begin{table*}[!htbp]
\centering
\caption{Complete Efficiency Metrics (SWE-bench Verified). Columns: $\mu_{\text{tool}}$ mean tool calls per trajectory; $\mu_{\text{LLM}}$ mean LLM invocations per trajectory; $\mu_{\text{bash}}$ mean bash invocations per trajectory; $P_{\text{pipe}}$ mean pipe (\texttt{|}) operators per bash command (a count, not a probability; values $> 1$ indicate multi-pipe chains); \textbf{Loc.} total localization-class commands across all trajectories; \textbf{Loc.\%} fraction of bash commands classified as localization. Qwen3-8B was run on SWE-Agent + SWE-bench (n=500 trajectories, included in the failure-mode partition Table~\ref{tab:failure_sweagent_combined}) but is excluded from per-tool aggregate tables due to its shorter average trajectories (7--10 tool calls vs.\ 20--40 for other models), which yield insufficient per-tool-type samples for reliable aggregation.}
\label{tab:app_swebench_efficiency}
\begin{tabular}{lcccccc}
\toprule
\textbf{Agent} & $\mu_{\text{tool}}$ & $\mu_{\text{LLM}}$ & $\mu_{\text{bash}}$ & $P_{\text{pipe}}$ & \textbf{Loc.} & \textbf{Loc.\%} \\
\midrule
GPT-5 & 30.71 & 30.71 & 15.36 & 0.22 & 2558 & 33.3 \\
GPT-5-mini & 32.16 & 32.16 & 16.84 & 0.24 & 3493 & 41.5 \\
Qwen3-235B & 42.53 & 42.53 & 15.25 & 0.20 & 1682 & 22.1 \\
Qwen3-Coder-30B & 50.45 & 50.45 & 27.44 & 0.42 & 2815 & 20.5 \\
Qwen3-32B & 48.59 & 48.59 & 13.62 & 0.23 & 2469 & 36.2 \\
Qwen3-Coder-480B & 50.30 & 50.30 & 24.80 & 0.36 & 2900 & 23.4 \\
\bottomrule
\end{tabular}
\end{table*}

\subsubsection{Token Consumption by Tool}

Table~\ref{tab:app_swebench_tokens} presents complete token consumption statistics for all tools across all models on SWE-bench Verified. Bash output tokens exhibit the highest variance across models. The view operation shows consistent input requirements (25.7--29.0 tokens) but substantial output variance (448--1682 tokens), indicating models inspect similar file regions but receive different context lengths based on file structure. The str\_replace operation reveals model-specific editing patterns: Qwen models require higher input tokens (190--616) compared to GPT models (232--327), suggesting more verbose edit specifications, while output tokens remain consistent (96--457), reflecting similar confirmation message lengths.

\begin{table*}[!htbp]
\centering
\caption{Complete Token Consumption (SWE-bench Verified). Qwen3-8B excluded from per-tool aggregates; see Table~\ref{tab:app_swebench_efficiency} caption.}
\label{tab:app_swebench_tokens}
\begin{tabular}{llrrrrrr}
\toprule
\textbf{Tool} & & \textbf{GPT-5} & \textbf{GPT-5-mini} & \textbf{Q-235B} & \textbf{Q-30B} & \textbf{Q-32B} & \textbf{Q-480B} \\
\midrule
\multirow{2}{*}{bash} & In & 76.0 & 96.8 & 41.8 & 87.0 & 44.7 & 54.2 \\
 & Out & 5210 & 1497 & 906 & 315 & 1301 & 593 \\
\midrule
\multirow{2}{*}{view} & In & 25.7 & 27.1 & 27.4 & 26.6 & 29.0 & 27.2 \\
 & Out & 1682 & 1406 & 902 & 776 & 448 & 715 \\
\midrule
\multirow{2}{*}{str\_replace} & In & 322 & 327 & 560 & 616 & 190 & 555 \\
 & Out & 316 & 379 & 418 & 457 & 132 & 432 \\
\midrule
\multirow{2}{*}{create} & In & 253 & 228 & 258 & 721 & 235 & 570 \\
 & Out & 15 & 16 & 18 & 15 & 20 & 16 \\
\midrule
\multirow{2}{*}{insert} & In & 149 & 111 & 132 & 61 & 103 & --- \\
 & Out & 231 & 244 & 261 & 182 & 234 & --- \\
\midrule
\multirow{2}{*}{undo\_edit} & In & 17.0 & 17.3 & 17.8 & --- & 20.4 & 17.1 \\
 & Out & 14.3 & 15.0 & 15.0 & --- & 15.7 & 14.4 \\
\midrule
\multirow{2}{*}{submit} & In & 5.3 & 5.3 & 5.3 & 5.3 & 5.3 & 5.3 \\
 & Out & 1864 & 351 & 2407 & 2583 & 614 & 1318 \\
\bottomrule
\end{tabular}
\end{table*}

\subsubsection{Tool Success Rates}

Table~\ref{tab:app_swebench_success} presents complete tool success rates for all models on SWE-bench Verified. The most striking anomaly is Qwen3-235B's submit success rate of only 26.5\% across 3557 attempts, compared to 99.9\% for GPT-5 and 100\% for Qwen3-Coder-30B, indicating systematic issues with patch formatting or validation logic in this configuration. Qwen3-32B also shows degraded performance across several file-manipulation operations: \texttt{create} drops to 78.9\%, \texttt{str\_replace} to 92.6\%, and \texttt{undo\_edit} to 71.7\%. In contrast, \texttt{view} maintains near-perfect reliability ($>$99\%) across all models, confirming file inspection as the most robust tool in the SWE-Agent toolkit. These operational failures are not captured by the stage-wise trajectory features but remain important for practical deployment.

\begin{table*}[!htbp]
\centering
\caption{Complete Tool Success Rates (SWE-bench Verified). Qwen3-8B excluded from per-tool aggregates; see Table~\ref{tab:app_swebench_efficiency} caption.}
\label{tab:app_swebench_success}
\begin{tabular}{lcccccc}
\toprule
\textbf{Tool} & \textbf{GPT-5} & \textbf{GPT-5-mini} & \textbf{Q-235B} & \textbf{Q-30B} & \textbf{Q-32B} & \textbf{Q-480B} \\
\midrule
bash & 96.2\% & 96.4\% & 96.3\% & 97.3\% & 94.5\% & 97.7\% \\
 & (7213) & (7953) & (7149) & (13225) & (6417) & (11905) \\
view & 99.7\% & 99.9\% & 99.9\% & 99.8\% & 99.2\% & 99.7\% \\
 & (4549) & (4200) & (4843) & (5970) & (7521) & (6165) \\
str\_replace & 98.8\% & 98.9\% & 97.2\% & 99.5\% & 92.6\% & 98.4\% \\
 & (1467) & (1656) & (4665) & (2086) & (8137) & (2748) \\
create & 98.8\% & 98.9\% & 92.7\% & 99.3\% & 78.9\% & 97.2\% \\
 & (659) & (264) & (1035) & (2901) & (1816) & (2493) \\
insert & 100\% & 100\% & 100\% & 100\% & 92.6\% & --- \\
 & (94) & (23) & (12) & (1) & (122) &  \\
undo\_edit & 100\% & 100\% & 100\% & --- & 71.7\% & 100\% \\
 & (67) & (15) & (1) &  & (46) & (2) \\
submit & 99.9\% & 99.9\% & 26.5\% & 100\% & 93.2\% & 99.9\% \\
 & (1305) & (1897) & (3557) & (1042) & (236) & (1838) \\
\bottomrule
\end{tabular}
\end{table*}

\subsection{Complete Model Results: PolyBench Verified}

\subsubsection{Efficiency Metrics}

Table~\ref{tab:app_polybench_efficiency} presents complete efficiency metrics for all seven models evaluated on PolyBench Verified. Cross-benchmark comparison reveals that multilingual tasks increase complexity: GPT-5 requires 25\% more tool invocations on PolyBench (38.38) compared to SWE-bench (30.71). The Qwen3-8B anomaly stands out with unusually low tool usage (24.89) and minimal localization (4.1\%), suggesting early termination or limited exploration capability rather than efficient problem-solving. Qwen3-32B exhibits the highest localization percentage (71.6\%), indicating extensive code navigation before editing---a pattern that correlates with its degraded tool success rates. Pipe usage increases substantially on PolyBench, with Qwen3-32B reaching 0.64 pipes per bash call, reflecting the need for complex command chaining in multilingual codebases with diverse build systems and directory structures.

\begin{table*}[!htbp]
\centering
\caption{Complete Efficiency Metrics (PolyBench Verified)}
\label{tab:app_polybench_efficiency}
\begin{tabular}{lcccccc}
\toprule
\textbf{Agent} & $\mu_{\text{tool}}$ & $\mu_{\text{LLM}}$ & $\mu_{\text{bash}}$ & $P_{\text{pipe}}$ & \textbf{Loc.} & \textbf{Loc.\%} \\
\midrule
GPT-5 & 38.38 & 38.38 & 18.06 & 0.34 & 2204 & 35.6 \\
GPT-5-mini & 34.62 & 34.62 & 18.89 & 0.37 & 2447 & 44.4 \\
Qwen3-235B & 43.12 & 43.12 & 18.26 & 0.41 & 2081 & 32.3 \\
Qwen3-32B & 48.45 & 48.45 & 30.26 & 0.64 & 7644 & 71.6 \\
Qwen3-8B & 24.89 & 24.89 & 24.89 & 0.02 & 362 & 4.1 \\
Qwen3-Coder-30B & 46.05 & 46.05 & 23.66 & 0.57 & 2149 & 25.7 \\
Qwen3-Coder-480B & 50.21 & 50.21 & 24.53 & 0.53 & 2435 & 28.1 \\
\bottomrule
\end{tabular}
\end{table*}

\subsubsection{Token Consumption by Tool}

Table~\ref{tab:app_polybench_tokens} presents complete token consumption statistics for all tools across all models on PolyBench Verified. Bash output tokens increase dramatically on PolyBench: GPT-5-mini reaches 21460 tokens (vs 1497 on SWE-bench), a 14$\times$ increase reflecting verbose compilation errors and test output in Java/TypeScript projects. The str\_replace output tokens for GPT-5 reach 1396 on PolyBench versus 316 on SWE-bench, a 4.4$\times$ increase indicating more complex edit contexts in multilingual codebases. File creation (create) requires substantially more input tokens on PolyBench (283--647 tokens) compared to SWE-bench (228--721 tokens), reflecting boilerplate requirements and import statements in statically-typed languages. The insert operation shows extreme variance for GPT-5-mini (2537 output tokens), suggesting verbose feedback when inserting code into Java/TypeScript files with complex type annotations.

\begin{table*}[!htbp]
\centering
\caption{Complete Token Consumption (PolyBench Verified)}
\setlength{\tabcolsep}{5pt} 
\label{tab:app_polybench_tokens}
\begin{tabular}{llrrrrrrr}
\toprule
\textbf{Tool} & & \textbf{GPT-5} & \textbf{GPT-5-mini} & \textbf{Q-235B} & \textbf{Q-32B} & \textbf{Q-8B} & \textbf{Q-30B} & \textbf{Q-480B} \\
\midrule
\multirow{2}{*}{bash} & In & 88.7 & 102.0 & 45.0 & 64.0 & 51.7 & 53.5 & 49.2 \\
 & Out & 12297 & 21460 & 3983 & 4344 & 968 & 2472 & 1539 \\
\midrule
\multirow{2}{*}{view} & In & 26.8 & 30.1 & 32.1 & 30.8 & --- & 28.3 & 29.0 \\
 & Out & 1800 & 1501 & 1012 & 711 & --- & 1244 & 1068 \\
\midrule
\multirow{2}{*}{str\_replace} & In & 306 & 314 & 438 & 190 & --- & 597 & 565 \\
 & Out & 1396 & 1313 & 611 & 130 & --- & 544 & 555 \\
\midrule
\multirow{2}{*}{create} & In & 336 & 341 & 293 & 283 & --- & 647 & 625 \\
 & Out & 62 & 97 & 38 & 16 & --- & 143 & 189 \\
\midrule
\multirow{2}{*}{insert} & In & 161 & 220 & 84 & 67 & --- & 352 & --- \\
 & Out & 510 & 2537 & 161 & 197 & --- & 518 & --- \\
\midrule
\multirow{2}{*}{undo\_edit} & In & 19.3 & 17.0 & --- & 19.7 & --- & --- & --- \\
 & Out & 16.7 & 14.2 & --- & 13.9 & --- & --- & --- \\
\midrule
\multirow{2}{*}{submit} & In & 5.3 & 5.3 & 5.3 & 5.3 & --- & 5.3 & 5.3 \\
 & Out & 3699 & 2882 & 6411 & 8258 & --- & 4961 & 12060 \\
\bottomrule
\end{tabular}
\end{table*}

\subsubsection{Tool Success Rates}

Table~\ref{tab:app_polybench_success} presents complete tool success rates for all models on PolyBench Verified. Cross-benchmark degradation is evident: Qwen3-32B's create success drops from 78.9\% (SWE-bench) to 58.7\% (PolyBench), and str\_replace falls from 92.6\% to 83.5\%. The undo\_edit operation shows severe degradation for Qwen3-32B, dropping from 71.7\% to 43.8\%, indicating that edit state management becomes increasingly problematic in multilingual contexts. Qwen3-235B's submit success improves from 26.5\% (SWE-bench) to 65.3\% (PolyBench), though this remains substantially below other models, suggesting partial but incomplete resolution of patch formatting issues. Qwen3-8B shows complete absence of editor tool usage (all marked as ---), confirming that this model fails to engage with file manipulation operations and relies exclusively on bash commands. The GPT-5 family maintains robust performance across benchmarks, with view operations achieving 100\% success and str\_replace remaining above 98.5\%.

\begin{table*}[!htbp]
\centering
\caption{Complete Tool Success Rates (PolyBench Verified)}
\label{tab:app_polybench_success}
\begin{tabular}{lccccccc}
\toprule
\textbf{Tool} & \textbf{GPT-5} & \textbf{GPT-5-mini} & \textbf{Q-235B} & \textbf{Q-32B} & \textbf{Q-8B} & \textbf{Q-30B} & \textbf{Q-480B} \\
\midrule
bash & 96.2\% & 95.9\% & 94.5\% & 92.9\% & 82.8\% & 95.9\% & 96.3\% \\
 & (5921) & (5266) & (6152) & (10514) & (8786) & (8043) & (8368) \\
view & 100\% & 99.9\% & 99.8\% & 96.6\% & --- & 99.9\% & 99.9\% \\
 & (4239) & (2846) & (4177) & (2993) &  & (4676) & (5391) \\
str\_replace & 99.4\% & 99.8\% & 98.2\% & 83.5\% & --- & 99.9\% & 99.2\% \\
 & (1757) & (1177) & (2946) & (3032) &  & (1381) & (1957) \\
create & 99.1\% & 99.6\% & 92.4\% & 58.7\% & --- & 99.8\% & 98.2\% \\
 & (450) & (233) & (1106) & (387) &  & (1545) & (1448) \\
insert & 100\% & 100\% & 100\% & 96.9\% & --- & 100\% & --- \\
 & (109) & (19) & (38) & (64) &  & (1) &  \\
undo\_edit & 100\% & 100\% & --- & 43.8\% & --- & --- & --- \\
 & (44) & (20) &  & (16) &  &  &  \\
submit & 99.8\% & 98.1\% & 65.3\% & 94.7\% & --- & 100\% & 100\% \\
 & (646) & (521) & (801) & (95) &  & (610) & (561) \\
\bottomrule
\end{tabular}
\end{table*}

\subsection{OpenHands Results}

\subsubsection{SWE-bench Verified}

Table~\ref{tab:app_openhands_swebench_efficiency} presents efficiency metrics for OpenHands on SWE-bench Verified. OpenHands demonstrates distinct behavioral patterns compared to SWE-agent: lower average tool usage (13.98--45.71 vs 30.71--50.45) but higher localization command percentages, with Qwen3-32B reaching 97.6\% localization. The Qwen3-Coder-480B configuration requires the most tool invocations (45.71) while maintaining moderate localization (61.7\%), suggesting a more balanced exploration-exploitation strategy. Pipe usage varies substantially, with Qwen3-235B showing anomalously high usage (1.10 pipes per bash call), indicating command chaining patterns with multiple pipes per invocation.

\begin{table*}[!htbp]
\centering
\caption{OpenHands Efficiency Metrics (SWE-bench Verified)}
\label{tab:app_openhands_swebench_efficiency}
\begin{tabular}{lcccccc}
\toprule
\textbf{Agent} & $\mu_{\text{tool}}$ & $\mu_{\text{LLM}}$ & $\mu_{\text{bash}}$ & $P_{\text{pipe}}$ & \textbf{Loc.} & \textbf{Loc.\%} \\
\midrule
GPT-5 & 24.18 & 24.18 & 18.84 & 0.56 & 7048 & 74.8 \\
GPT-5-mini & 20.06 & 20.06 & 15.15 & 0.49 & 5627 & 74.3 \\
Qwen3-235B & 26.26 & 26.26 & 9.65 & 1.10 & 3650 & 75.7 \\
Qwen3-32B & 13.98 & 13.98 & 4.49 & 0.42 & 2192 & 97.6 \\
Qwen3-8B & 26.55 & 26.55 & 8.97 & 0.45 & 4188 & 93.4 \\
Qwen3-Coder-480B & 45.71 & 45.71 & 20.61 & 0.74 & 6354 & 61.7 \\
Qwen3-Coder-30B & 37.48 & 37.48 & 18.92 & 0.76 & 5579 & 59.0 \\
\bottomrule
\end{tabular}
\end{table*}

Table~\ref{tab:app_openhands_swebench_tokens} presents token consumption for OpenHands tools on SWE-bench. The fsWrite operation shows substantial variance in input tokens (230--694), reflecting different code generation verbosity across models. Bash output tokens range from 195 (Qwen3-Coder-480B) to 4503 (Qwen3-32B), a 23$\times$ difference indicating divergent command strategies.

\begin{table*}[!htbp]
\centering
\caption{OpenHands Token Consumption (SWE-bench Verified)}
\label{tab:app_openhands_swebench_tokens}
\begin{tabular}{llrrrrrrr}
\toprule
\textbf{Tool} & & \textbf{GPT-5} & \textbf{GPT-5-mini} & \textbf{Q-235B} & \textbf{Q-32B} & \textbf{Q-8B} & \textbf{Q-480B} & \textbf{Q-30B} \\
\midrule
\multirow{2}{*}{executeBash} & In & 31.6 & 34.3 & 28.1 & 33.0 & 22.9 & 31.6 & 41.0 \\
 & Out & 1046 & 1555 & 549 & 4503 & 1962 & 195 & 208 \\
\midrule
\multirow{2}{*}{fsRead} & In & 18.7 & 17.4 & 18.4 & 18.7 & 18.7 & 18.2 & 18.5 \\
 & Out & 1694 & 1610 & 1305 & 510 & 500 & 695 & 753 \\
\midrule
\multirow{2}{*}{fsWrite} & In & 277 & 417 & 324 & 230 & 374 & 593 & 694 \\
 & Out & 332 & 250 & 267 & 141 & 131 & 206 & 244 \\
\midrule
\multirow{2}{*}{system} & In & 3104 & 3104 & 3528 & 3528 & 3528 & 3528 & 3528 \\
 & Out & 5.8 & 5.8 & 5.8 & 5.8 & 5.8 & 5.8 & 5.8 \\
\midrule
\multirow{2}{*}{think} & In & 481 & 296 & 252 & 537 & 391 & 479 & 634 \\
 & Out & 7.3 & 7.3 & 7.3 & 7.3 & 7.3 & 7.3 & 7.3 \\
\bottomrule
\end{tabular}
\end{table*}

Table~\ref{tab:app_openhands_swebench_success} presents tool success rates for OpenHands on SWE-bench. The fsWrite operation exhibits the most variance, ranging from 29.0\% (Qwen3-8B) to 97.7\% (Qwen3-Coder-480B). This 3.4$\times$ difference highlights model-specific challenges with file creation and modification. The fsRead operation shows degraded performance for GPT-5-mini (63.2\%) and Qwen3-32B (59.8\%), suggesting file path resolution issues. System and think tools maintain 100\% success across all models, confirming their reliability as non-destructive operations.

\begin{table*}[!htbp]
\centering
\caption{OpenHands Tool Success Rates (SWE-bench Verified)}
\label{tab:app_openhands_swebench_success}
\begin{tabular}{lccccccc}
\toprule
\textbf{Tool} & \textbf{GPT-5} & \textbf{GPT-5-mini} & \textbf{Q-235B} & \textbf{Q-32B} & \textbf{Q-8B} & \textbf{Q-480B} & \textbf{Q-30B} \\
\midrule
executeBash & 95.4\% & 96.3\% & 91.5\% & 98.0\% & 97.0\% & 94.9\% & 95.1\% \\
 & (9418) & (7575) & (4823) & (2246) & (4483) & (10304) & (9462) \\
fsRead & 87.4\% & 63.2\% & 98.2\% & 59.8\% & 95.5\% & 97.8\% & 98.2\% \\
 & (952) & (960) & (2953) & (3260) & (4159) & (7930) & (5345) \\
fsWrite & 89.0\% & 74.1\% & 76.1\% & 44.5\% & 29.0\% & 97.7\% & 95.6\% \\
 & (1168) & (953) & (927) & (483) & (1143) & (2803) & (1817) \\
system & 100\% & 100\% & 100\% & 100\% & 100\% & 100\% & 100\% \\
 & (498) & (492) & (500) & (500) & (500) & (500) & (500) \\
think & 100\% & 100\% & 100\% & 100\% & 100\% & 100\% & 100\% \\
 & (55) & (20) & (3929) & (501) & (1964) & (1317) & (1618) \\
\bottomrule
\end{tabular}
\end{table*}

\subsubsection{PolyBench Verified}

Table~\ref{tab:app_openhands_polybench_efficiency} presents efficiency metrics for OpenHands on PolyBench Verified. Cross-benchmark comparison reveals increased complexity: GPT-5 requires 20\% more tool invocations on PolyBench (29.03) compared to SWE-bench (24.18). Qwen3-Coder-480B maintains the highest tool usage (47.45) with moderate localization (66.2\%). The Qwen3-Coder-30B configuration shows elevated pipe usage (0.88 pipes per bash call), reflecting complex command chaining requirements in multilingual codebases.

\begin{table*}[!htbp]
\centering
\caption{OpenHands Efficiency Metrics (PolyBench Verified)}
\label{tab:app_openhands_polybench_efficiency}
\begin{tabular}{lcccccc}
\toprule
\textbf{Agent} & $\mu_{\text{tool}}$ & $\mu_{\text{LLM}}$ & $\mu_{\text{bash}}$ & $P_{\text{pipe}}$ & \textbf{Loc.} & \textbf{Loc.\%} \\
\midrule
GPT-5 & 29.03 & 29.03 & 20.18 & 0.56 & 5884 & 76.3 \\
GPT-5-mini & 25.45 & 25.45 & 19.59 & 0.51 & 5449 & 72.8 \\
Qwen3-235B & 30.40 & 30.40 & 14.68 & 0.84 & 4511 & 80.5 \\
Qwen3-32B & 18.03 & 18.03 & 8.83 & 0.93 & 2786 & 82.6 \\
Qwen3-8B & 25.43 & 25.43 & 10.98 & 0.30 & 3463 & 82.6 \\
Qwen3-Coder-480B & 47.45 & 47.45 & 19.58 & 0.66 & 4956 & 66.2 \\
Qwen3-Coder-30B & 37.80 & 37.80 & 17.38 & 0.88 & 4421 & 66.6 \\
\bottomrule
\end{tabular}
\end{table*}

Table~\ref{tab:app_openhands_polybench_tokens} presents token consumption for OpenHands on PolyBench. Bash output tokens increase substantially: GPT-5 reaches 2329 tokens (vs 1046 on SWE-bench), while Qwen3-8B shows extreme output (6628 tokens), indicating verbose error messages in multilingual compilation. The fsWrite input tokens remain relatively stable across benchmarks, suggesting consistent code generation patterns regardless of target language.

\begin{table*}[!htbp]
\centering
\caption{OpenHands Token Consumption (PolyBench Verified)}
\setlength{\tabcolsep}{5pt} 
\label{tab:app_openhands_polybench_tokens}
\begin{tabular}{llrrrrrrr}
\toprule
\textbf{Tool} & & \textbf{GPT-5} & \textbf{GPT-5-mini} & \textbf{Q-235B} & \textbf{Q-32B} & \textbf{Q-8B} & \textbf{Q-480B} & \textbf{Q-30B} \\
\midrule
\multirow{2}{*}{executeBash} & In & 40.3 & 38.6 & 30.8 & 41.5 & 26.5 & 26.0 & 35.3 \\
 & Out & 2329 & 3086 & 1016 & 1290 & 6628 & 524 & 623 \\
\midrule
\multirow{2}{*}{fsRead} & In & 23.5 & 24.4 & 24.2 & 23.9 & 25.0 & 23.0 & 23.1 \\
 & Out & 1516 & 1122 & 1348 & 575 & 498 & 690 & 823 \\
\midrule
\multirow{2}{*}{fsWrite} & In & 311 & 383 & 363 & 279 & 472 & 572 & 606 \\
 & Out & 308 & 253 & 240 & 154 & 124 & 270 & 285 \\
\midrule
\multirow{2}{*}{system} & In & 3104 & 3104 & 3528 & 3528 & 3528 & 3528 & 3528 \\
 & Out & 5.8 & 5.8 & 5.8 & 5.8 & 5.8 & 5.8 & 5.8 \\
\midrule
\multirow{2}{*}{think} & In & 688 & 349 & 281 & 138 & 455 & 523 & 701 \\
 & Out & 7.3 & 7.3 & 7.3 & 7.4 & 7.3 & 7.3 & 7.3 \\
\bottomrule
\end{tabular}
\end{table*}

Table~\ref{tab:app_openhands_polybench_success} presents tool success rates for OpenHands on PolyBench. The fsWrite operation shows degraded performance across most models: GPT-5 drops from 89.0\% (SWE-bench) to 66.7\% (PolyBench), and Qwen3-8B falls from 29.0\% to 20.6\%. This cross-benchmark degradation reflects increased complexity in multilingual file manipulation. The fsRead operation also degrades, with GPT-5-mini dropping from 63.2\% to 61.0\%.

\begin{table*}[!htbp]
\centering
\caption{OpenHands Tool Success Rates (PolyBench Verified)}
\label{tab:app_openhands_polybench_success}
\begin{tabular}{lccccccc}
\toprule
\textbf{Tool} & \textbf{GPT-5} & \textbf{GPT-5-mini} & \textbf{Q-235B} & \textbf{Q-32B} & \textbf{Q-8B} & \textbf{Q-480B} & \textbf{Q-30B} \\
\midrule
executeBash & 98.2\% & 97.8\% & 96.2\% & 98.6\% & 98.1\% & 95.3\% & 96.0\% \\
 & (7710) & (7483) & (5607) & (3374) & (4195) & (7481) & (6641) \\
fsRead & 89.2\% & 61.0\% & 97.3\% & 66.3\% & 90.0\% & 93.2\% & 97.7\% \\
 & (1217) & (677) & (2883) & (2401) & (2465) & (7016) & (4621) \\
fsWrite & 66.7\% & 55.4\% & 54.5\% & 39.8\% & 20.6\% & 90.5\% & 77.8\% \\
 & (1722) & (1140) & (1264) & (530) & (880) & (2087) & (1540) \\
system & 100\% & 100\% & 100\% & 100\% & 100\% & 100\% & 100\% \\
 & (365) & (375) & (382) & (382) & (380) & (382) & (382) \\
think & 100\% & 100\% & 100\% & 100\% & 100\% & 100\% & 100\% \\
 & (75) & (47) & (1478) & (200) & (824) & (1158) & (1255) \\
\bottomrule
\end{tabular}
\end{table*}

\subsection{LiveSWEAgent Results}

\subsubsection{SWE-bench Verified}

Table~\ref{tab:app_liveswe_swebench_efficiency} presents efficiency metrics for LiveSWEAgent on SWE-bench Verified. LiveSWEAgent operates exclusively through bash commands, with tool usage equal to bash usage across all models. The GPT-5 family shows efficient operation (13.45--14.67 calls per task), while Qwen models require substantially more invocations (42.38--50.56). Localization percentages range from 50.9\% (Qwen3-Coder-30B) to 66.4\% (Qwen3-Coder-480B), indicating moderate exploration overhead. Pipe usage is notably higher for GPT-5 models (0.74--0.75 pipes per bash call) compared to Qwen3-235B (0.26), suggesting different command composition strategies.

\begin{table*}[!htbp]
\centering
\caption{LiveSWEAgent Efficiency Metrics (SWE-bench Verified). Qwen3-8B and Qwen3-32B excluded due to OpenRouter API errors (Section~\ref{sec:experiments}, coverage note).}
\label{tab:app_liveswe_swebench_efficiency}
\begin{tabular}{lcccccc}
\toprule
\textbf{Agent} & $\mu_{\text{tool}}$ & $\mu_{\text{LLM}}$ & $\mu_{\text{bash}}$ & $P_{\text{pipe}}$ & \textbf{Loc.} & \textbf{Loc.\%} \\
\midrule
GPT-5 & 13.45 & 13.45 & 13.45 & 0.75 & 4224 & 62.8 \\
GPT-5-mini & 14.67 & 14.67 & 14.67 & 0.74 & 4545 & 61.9 \\
Qwen3-235B & 43.76 & 43.76 & 43.76 & 0.26 & 14447 & 66.0 \\
Qwen3-Coder-30B & 42.38 & 42.38 & 42.38 & 0.44 & 10782 & 50.9 \\
Qwen3-Coder-480B & 50.56 & 50.56 & 50.56 & 0.21 & 16793 & 66.4 \\
\bottomrule
\end{tabular}
\end{table*}

Table~\ref{tab:app_liveswe_swebench_tokens} presents token consumption for LiveSWEAgent on SWE-bench. Bash input tokens range from 79.8 (Qwen3-Coder-30B) to 158.8 (GPT-5), reflecting different command verbosity. Output tokens show extreme variance: Qwen3-235B generates 14688 tokens per command on average, compared to 152 tokens for Qwen3-Coder-480B---a 97$\times$ difference indicating fundamentally different execution patterns, likely due to verbose error handling or extensive test output.

\begin{table*}[!htbp]
\centering
\caption{LiveSWEAgent Token Consumption (SWE-bench Verified). Qwen3-8B and Qwen3-32B excluded due to OpenRouter API errors.}
\label{tab:app_liveswe_swebench_tokens}
\begin{tabular}{llrrrrr}
\toprule
\textbf{Tool} & & \textbf{GPT-5} & \textbf{GPT-5-mini} & \textbf{Q-235B} & \textbf{Q-30B} & \textbf{Q-480B} \\
\midrule
\multirow{2}{*}{executeBash} & In & 158.8 & 96.7 & 122.3 & 79.8 & 97.5 \\
 & Out & 528 & 430 & 14688 & 817 & 152 \\
\bottomrule
\end{tabular}
\end{table*}

Table~\ref{tab:app_liveswe_swebench_success} presents tool success rates for LiveSWEAgent on SWE-bench. Bash success rates range from 92.9\% (Qwen3-Coder-30B) to 98.2\% (Qwen3-Coder-480B). The GPT-5 family maintains consistent performance (94.3--95.8\%), while Qwen models show more variance. The high success rates across all models confirm bash command execution as a reliable operation, though the variance suggests model-specific command formatting issues.

\begin{table*}[!htbp]
\centering
\caption{LiveSWEAgent Tool Success Rates (SWE-bench Verified). Qwen3-8B and Qwen3-32B excluded due to OpenRouter API errors.}
\label{tab:app_liveswe_swebench_success}
\begin{tabular}{lccccc}
\toprule
\textbf{Tool} & \textbf{GPT-5} & \textbf{GPT-5-mini} & \textbf{Q-235B} & \textbf{Q-30B} & \textbf{Q-480B} \\
\midrule
executeBash & 94.3\% & 95.8\% & 97.8\% & 92.9\% & 98.2\% \\
 & (5434) & (6010) & (20880) & (17701) & (24474) \\
\bottomrule
\end{tabular}
\end{table*}

\subsubsection{PolyBench Verified}

Table~\ref{tab:app_liveswe_polybench_efficiency} presents efficiency metrics for LiveSWEAgent on PolyBench Verified. Cross-benchmark comparison shows mixed patterns: GPT-5 requires slightly more invocations on PolyBench (15.23 vs 13.45), while Qwen3-235B shows reduced usage (25.10 vs 43.76). Localization percentages increase for most models on PolyBench, with Qwen3-Coder-480B reaching 68.9\%, reflecting additional navigation requirements in multilingual codebases. Pipe usage increases substantially for GPT-5 (0.89 vs 0.75 pipes per bash call), indicating more complex command chaining for multilingual tasks.

\begin{table*}[!htbp]
\centering
\caption{LiveSWEAgent Efficiency Metrics (PolyBench Verified). Qwen3-8B and Qwen3-32B excluded due to OpenRouter API errors.}
\label{tab:app_liveswe_polybench_efficiency}
\begin{tabular}{lcccccc}
\toprule
\textbf{Agent} & $\mu_{\text{tool}}$ & $\mu_{\text{LLM}}$ & $\mu_{\text{bash}}$ & $P_{\text{pipe}}$ & \textbf{Loc.} & \textbf{Loc.\%} \\
\midrule
GPT-5 & 15.23 & 15.23 & 15.23 & 0.89 & 3906 & 67.1 \\
GPT-5-mini & 14.28 & 14.28 & 14.28 & 0.86 & 3674 & 67.3 \\
Qwen3-235B & 25.10 & 25.10 & 25.10 & 0.31 & 6299 & 65.7 \\
Qwen3-Coder-30B & 42.33 & 42.33 & 42.33 & 0.46 & 7703 & 47.6 \\
Qwen3-Coder-480B & 46.78 & 46.78 & 46.78 & 0.26 & 12311 & 68.9 \\
\bottomrule
\end{tabular}
\end{table*}

Table~\ref{tab:app_liveswe_polybench_tokens} presents token consumption for LiveSWEAgent on PolyBench. Bash input tokens remain relatively stable across benchmarks, with GPT-5 at 148.8 tokens (vs 158.8 on SWE-bench). Output tokens decrease substantially for Qwen3-235B (230 vs 14688), suggesting more efficient command execution in multilingual contexts, possibly due to different error handling patterns or reduced test verbosity.

\begin{table*}[!htbp]
\centering
\caption{LiveSWEAgent Token Consumption (PolyBench Verified). Qwen3-8B and Qwen3-32B excluded due to OpenRouter API errors.}
\label{tab:app_liveswe_polybench_tokens}
\begin{tabular}{llrrrrr}
\toprule
\textbf{Tool} & & \textbf{GPT-5} & \textbf{GPT-5-mini} & \textbf{Q-235B} & \textbf{Q-30B} & \textbf{Q-480B} \\
\midrule
\multirow{2}{*}{executeBash} & In & 148.8 & 95.5 & 143.6 & 89.3 & 101.1 \\
 & Out & 501 & 466 & 230 & 131 & 191 \\
\bottomrule
\end{tabular}
\end{table*}

Table~\ref{tab:app_liveswe_polybench_success} presents tool success rates for LiveSWEAgent on PolyBench. Success rates remain high across all models (93.3--97.9\%), with Qwen3-Coder-30B achieving the highest rate (97.9\%) despite moderate efficiency. GPT-5-mini shows slight degradation (93.3\% vs 95.8\% on SWE-bench), while Qwen3-Coder-480B maintains consistent performance (97.6\% vs 98.2\%). These results confirm bash-only operation as a robust strategy across benchmark types.

\begin{table*}[!htbp]
\centering
\caption{LiveSWEAgent Tool Success Rates (PolyBench Verified). Qwen3-8B and Qwen3-32B excluded due to OpenRouter API errors.}
\label{tab:app_liveswe_polybench_success}
\begin{tabular}{lccccc}
\toprule
\textbf{Tool} & \textbf{GPT-5} & \textbf{GPT-5-mini} & \textbf{Q-235B} & \textbf{Q-30B} & \textbf{Q-480B} \\
\midrule
executeBash & 94.5\% & 93.3\% & 95.3\% & 97.9\% & 97.6\% \\
 & (4653) & (4235) & (8657) & (14358) & (17184) \\
\bottomrule
\end{tabular}
\end{table*}

\subsection{Algorithm: Trajectory Feature Extraction}
\label{sec:appendix_algorithm}

\begin{algorithm}[h]
\caption{Trajectory Feature Extraction}
\label{alg:extraction}
\begin{algorithmic}[1]
\REQUIRE Trajectory $T = \{(a_t, o_t)\}_{t=1}^{n}$, Reference patch $P^*$, Tool mapping $\phi$
\ENSURE Feature vector $\mathbf{m} \in \mathbb{R}^6$
\STATE $\mathcal{F}^*, \mathcal{H}^* \gets \textsc{ParsePatch}(P^*)$
\STATE $\mathcal{F}_T, \mathcal{H}_T, \mathcal{H}_{\hat{P}} \gets \emptyset, \emptyset, \emptyset$
\FOR{$(a_t, o_t) \in T$}
    \IF{$\phi(a_t) = \texttt{view}$}
        \STATE $\mathcal{F}_T \gets \mathcal{F}_T \cup \textsc{ExtractFiles}(a_t)$
        \STATE $\mathcal{H}_T \gets \mathcal{H}_T \cup \textsc{ExtractFunctions}(o_t)$
    \ELSIF{$\phi(a_t) = \texttt{edit}$}
        \STATE $\mathcal{H}_{\hat{P}} \gets \mathcal{H}_{\hat{P}} \cup \textsc{ExtractEditTargets}(a_t)$
    \ENDIF
\ENDFOR
\STATE $\mathbf{m} \gets [P_s, R_s, P_r, R_r, P_e, R_e]$ using Eqs.~\ref{eq:search}--\ref{eq:edit}
\STATE \textbf{Return} $\mathbf{m}$
\end{algorithmic}
\end{algorithm}

\subsection{Trajectory Feature Extraction Methodology}

This section details the feature extraction pipeline used to compute the six trajectory metrics.

\subsubsection{Golden Context Definition}

We define \textbf{golden context} as the minimal code elements required to generate the correct patch:

\begin{enumerate}
\item \textbf{Golden Files:} Files modified in the ground-truth patch, extracted from diff headers (\texttt{diff --git a/... b/...})
\item \textbf{Golden Functions:} Functions containing modified lines, extracted from hunk headers (\texttt{@@ ... @@ function\_name})
\item \textbf{Golden Lines:} Line ranges modified in each file, extracted from hunk markers
\end{enumerate}

\subsubsection{Agent Action Extraction}

For each agent framework, we extract actions from trajectory logs:

\paragraph{SWE-Agent Actions:}
\begin{itemize}
\item \texttt{view}: File opened for inspection $\rightarrow$ files\_viewed
\item \texttt{str\_replace}: Edit operation $\rightarrow$ files\_edited
\item \texttt{bash}: Commands like \texttt{cat}, \texttt{grep} $\rightarrow$ files\_viewed
\end{itemize}

\paragraph{OpenHands Actions:}
\begin{itemize}
\item \texttt{fsRead}: File read operation $\rightarrow$ files\_viewed
\item \texttt{fsWrite}: File write operation $\rightarrow$ files\_edited
\item \texttt{executeBash}: Shell commands parsed for file references
\end{itemize}

\paragraph{LiveSWEAgent Actions:}
\begin{itemize}
\item \texttt{executeBash}: All operations via bash commands
\item Parse \texttt{cat}, \texttt{head}, \texttt{vim}, \texttt{sed} for files\_viewed
\item Parse \texttt{patch}, \texttt{sed -i}, file redirects for files\_edited
\end{itemize}

\subsubsection{Metric Computation}

Given golden context $G$ and agent actions $A$:

\begin{align}
\text{Search Precision} &= \frac{|G_{\text{files}} \cap A_{\text{viewed}}|}{|A_{\text{viewed}}|} \\
\text{Search Recall} &= \frac{|G_{\text{files}} \cap A_{\text{viewed}}|}{|G_{\text{files}}|} \\
\text{Read Precision} &= \frac{|G_{\text{funcs}} \cap A_{\text{funcs\_read}}|}{|A_{\text{funcs\_read}}|} \\
\text{Read Recall} &= \frac{|G_{\text{funcs}} \cap A_{\text{funcs\_read}}|}{|G_{\text{funcs}}|} \\
\text{Edit Precision} &= \frac{|G_{\text{funcs}} \cap A_{\text{funcs\_edited}}|}{|A_{\text{funcs\_edited}}|} \\
\text{Edit Recall} &= \frac{|G_{\text{funcs}} \cap A_{\text{funcs\_edited}}|}{|G_{\text{funcs}}|}
\end{align}

\subsection{Threshold Sensitivity for Failure-Mode Buckets}
\label{sec:appendix_threshold}

The failure-mode partition in Section~\ref{sec:failure_modes} uses a cascading 0.5 threshold on stage-wise recall. We re-bucket the same SWE-Agent + SWE-bench failures at $\tau \in \{0.3, 0.5, 0.7\}$ to test whether the qualitative findings depend on this choice. Lower $\tau$ relaxes the bar for ``reaching'' a stage; higher $\tau$ makes it stricter.

\begin{table}[!htbp]
\centering
\small
\setlength{\tabcolsep}{3pt}
\caption{Failure-mode distribution under varying recall thresholds $\tau$, SWE-Agent on SWE-bench Verified. Each row sums to 100\%. Bold entries mark the modal bucket per row. Edit-Quality remains the modal bucket for all five capable models at every threshold (53--72\%); Qwen3-32B and Qwen3-8B move between Search-modal and Edit-Quality-modal but stay in the upstream-failure regime (Search + Read $\geq$ 50\% at $\tau=0.7$).}
\label{tab:failure_threshold_sensitivity}
\begin{tabular}{lcrcccc}
\toprule
\textbf{Model} & $\tau$ & \textbf{$n$} & \textbf{Search} & \textbf{Read} & \textbf{E.-Tgt.} & \textbf{E.-Qual.} \\
\midrule
\multirow{3}{*}{GPT-5}            & 0.3 & 163 &  5.5 &  4.9 & 19.6 & \textbf{69.9} \\
                                  & 0.5 & 163 &  6.7 &  4.3 & 20.9 & \textbf{68.1} \\
                                  & 0.7 & 163 & 18.4 &  4.9 & 20.9 & \textbf{55.8} \\
\midrule
\multirow{3}{*}{GPT-5-mini}       & 0.3 & 162 &  6.8 &  7.4 & 14.2 & \textbf{71.6} \\
                                  & 0.5 & 162 &  8.6 &  8.0 & 14.2 & \textbf{69.1} \\
                                  & 0.7 & 162 & 16.7 &  8.6 & 16.0 & \textbf{58.6} \\
\midrule
\multirow{3}{*}{Qwen3-Coder-480B} & 0.3 & 187 &  7.0 &  7.5 & 15.0 & \textbf{70.6} \\
                                  & 0.5 & 187 &  9.1 & 10.7 & 13.4 & \textbf{66.8} \\
                                  & 0.7 & 187 & 18.7 & 15.0 & 10.7 & \textbf{55.6} \\
\midrule
\multirow{3}{*}{Qwen3-Coder-30B}  & 0.3 & 231 & 10.0 &  8.2 & 11.7 & \textbf{70.1} \\
                                  & 0.5 & 231 & 11.7 & 10.8 & 10.8 & \textbf{66.7} \\
                                  & 0.7 & 231 & 22.1 & 11.7 & 11.3 & \textbf{55.0} \\
\midrule
\multirow{3}{*}{Qwen3-235B}       & 0.3 & 249 & 15.7 &  6.0 & 10.8 & \textbf{67.5} \\
                                  & 0.5 & 249 & 17.7 &  7.2 & 10.4 & \textbf{64.7} \\
                                  & 0.7 & 249 & \textbf{28.1} &  6.0 & 12.4 & 53.4 \\
\midrule
\multirow{3}{*}{Qwen3-32B}        & 0.3 & 362 & 24.6 &  9.9 & 27.1 & \textbf{38.4} \\
                                  & 0.5 & 362 & 25.7 &  9.4 & \textbf{27.3} & 37.6 \\
                                  & 0.7 & 362 & \textbf{32.0} &  9.4 & 24.6 & 34.0 \\
\midrule
\multirow{3}{*}{Qwen3-8B}         & 0.3 & 253 & 39.1 &  9.9 &  2.4 & \textbf{48.6} \\
                                  & 0.5 & 253 & 41.1 & 11.1 &  2.4 & \textbf{45.5} \\
                                  & 0.7 & 253 & \textbf{47.8} &  9.1 &  2.8 & 40.3 \\
\midrule
\multicolumn{2}{l}{\textbf{Weighted avg.}} & 1607 &  &  &  &  \\
~~at $\tau=0.3$ &     &      & 17.6 &  8.0 & 15.0 & \textbf{59.4} \\
~~at $\tau=0.5$ &     &      & 19.3 &  9.0 & 14.8 & \textbf{56.9} \\
~~at $\tau=0.7$ &     &      & 28.0 &  9.3 & 14.5 & \textbf{48.2} \\
\bottomrule
\end{tabular}
\end{table}

\paragraph{Findings.} Three observations support the robustness of the Section~\ref{sec:failure_modes} claims. (1) Edit-Quality remains the modal failure mode for all five capable models (GPT-5, GPT-5-mini, Qwen3-Coder-480B/30B, Qwen3-235B) at every threshold; the rate spans 53--72\%, well above any other bucket. (2) The capable-vs-capability-bound split is preserved: Qwen3-8B and Qwen3-32B remain in the upstream-failure regime (Search + Read $\geq$ 50\% at $\tau$=0.7), while GPT-5-family models stay Edit-Quality-dominated. (3) The shift at $\tau$=0.7 is concentrated in Search at the expense of Edit-Quality (capable models lose 10--14 points of Edit-Quality, gain $\sim$10 points of Search), consistent with stricter bars demoting ``barely reached'' trajectories from downstream to upstream buckets---an effect on labels rather than on the underlying behavior. The qualitative conclusions of Section~\ref{sec:failure_modes} are stable.

\subsection{Reference-Tier Sensitivity: Tier-0 vs.\ Tier-1 Golden Context}
\label{sec:appendix_tier_sensitivity}

The golden context $\mathcal{G}$ used throughout the paper (Section~\ref{sec:approach}) is the strict set of files and functions modified by the reference patch (\emph{Tier-0}). A natural alternative is to expand the reference one hop along the call graph---the patched function plus its direct callers and callees and their containing files (\emph{Tier-1})---on the grounds that an agent that reads a caller of the buggy function is doing useful work even if the caller is not itself patched. We re-extract the golden context under Tier-1 and recompute every stage-wise precision/recall and failure-mode bucket on the same 16{,}758 trajectories. The script is included in the artifact (\texttt{scripts/recompute\_tier1\_breakdown.py}); the recomputed per-trajectory data is released as \texttt{failure\_mode\_breakdown\_tier1.csv} alongside the Tier-0 CSV.

\paragraph{Tier-1 expansion.} On average, Tier-1 reference sets contain 11.7$\times$ more files than Tier-0 (1 patched file $\rightarrow$ $\sim$12 files in the Tier-1 reference). The Tier-1 reference is therefore an \emph{upper bound} on what an agent could reasonably need to inspect; Tier-0 is a \emph{lower bound} (only what was actually changed).

\paragraph{Effect on stage-wise recall and failure-mode buckets.} Mean Search recall drops from 0.756 (Tier-0) to 0.292 (Tier-1) across all trajectories: agents rarely view every one-hop dependency, even on tasks they ultimately resolve. The cascading 0.5-threshold partition (Section~\ref{sec:failure_modes}) is sensitive to this drop: 4,379 trajectories that were assigned to downstream failure buckets under Tier-0 shift to the \emph{Search} bucket under Tier-1, because their Search recall falls below 0.5 once measured against the larger denominator. The Tier-1 partition therefore concentrates almost all failure mass at Search and loses the discrimination among Read, Edit-Target, and Edit-Quality that Section~\ref{sec:failure_modes} relies on.

\paragraph{What \emph{does} survive under Tier-1.}
We verified two consistency properties: (i) every trajectory's Tier-1 recall is $\leq$ its Tier-0 recall at every stage, as required by set inclusion (0 violations across 16{,}758 trajectories); and (ii) the model ranking by aggregate recall is identical under Tier-0 and Tier-1 (Spearman $\rho = 1.00$). The decomposition is therefore robust to the choice of reference tier in the relative sense---which model is more recall-bound than which---but the absolute stage-wise rates and the failure-mode partition depend on the tier.

\paragraph{Why we report Tier-0 in the main paper.} Two reasons. First, Tier-0 has an unambiguous semantics: a trajectory that fails to view a Tier-0 file failed to view a file the agent demonstrably needed to modify. Tier-1 includes files an agent \emph{might} have benefited from viewing but did not strictly need; treating those as required inflates the apparent failure rate at Search and obscures the very downstream failure modes (Edit-Target, Edit-Quality) that motivate this paper. Second, Tier-0 is benchmark-portable: any benchmark with reference patches yields a Tier-0 reference for free, whereas Tier-1 requires a working call-graph extractor per language. Reporting Tier-1 as a sensitivity check rather than as the primary reference preserves portability and discriminative power simultaneously.

\paragraph{Why an intermediate ``Tier-0.5'' is also too coarse.} A natural objection to the Tier-0 vs.\ Tier-1 dichotomy is that the right reference might lie between the two---e.g., the patched function plus all other functions in the same file. We tested this intermediate tier: the mean number of functions per gold file is 102, yielding a 66$\times$ expansion of the function-level reference set, larger than Tier-1's 11.7$\times$ file-level expansion. Under this intermediate tier, 94.5\% of trajectories collapse into the Read-failure bucket, exceeding the Search-failure concentration produced by Tier-1 itself. The intermediate tier is therefore not a useful middle ground: any reference broader than ``the functions actually modified'' becomes too permissive at the function granularity \codename{} operates on, because gold files typically contain many functions unrelated to the bug. This finding sharpens the design choice: function-level Tier-0 is the appropriate granularity for the search/read/edit decomposition, not because intermediate tiers were untested but because they fail the discrimination criterion empirically.

\subsection{Failure-Mode Distribution: Remaining (Scaffold, Benchmark) Cells}
\label{sec:appendix_failure_modes}

This section extends the failure-mode breakdown introduced in Section~\ref{sec:failure_modes} to the remaining five (scaffold, benchmark) cells. Every failed trajectory (Pass@1 = 0) is partitioned into one of four mutually exclusive, exhaustive buckets using a cascading 0.5 threshold on stage-wise recall: \textbf{Search} ($R_s < 0.5$), \textbf{Read} ($R_s \geq 0.5$, $R_r < 0.5$), \textbf{Edit-Target} ($R_r \geq 0.5$, $R_e < 0.5$), and \textbf{Edit-Quality} ($R_e \geq 0.5$). Rows sum to 100\%.

\begin{table}[!htbp]
\centering
\small
\setlength{\tabcolsep}{3pt}
\caption{OpenHands on SWE-bench Verified.}
\label{tab:failure_openhands_swebench}
\begin{tabular}{lrcccc}
\toprule
\textbf{Model} & \textbf{$n$} & \textbf{Search} & \textbf{Read} & \textbf{E.-Tgt.} & \textbf{E.-Qual.} \\
\midrule
GPT-5            & 158 & 11.4 &  3.2 & 17.7 & \textbf{67.7} \\
GPT-5-mini       & 204 & 10.3 &  4.4 & 24.5 & \textbf{60.8} \\
Qwen3-Coder-480B & 196 &  9.7 & 10.2 & 15.3 & \textbf{64.8} \\
Qwen3-Coder-30B  & 278 &  8.6 & 12.2 & 14.7 & \textbf{64.4} \\
Qwen3-235B       & 276 & 14.9 &  9.8 & 10.9 & \textbf{64.5} \\
Qwen3-32B        & 445 & \textbf{41.8} & 18.9 & 17.5 & 21.8 \\
Qwen3-8B         & 444 & \textbf{43.0} & 19.8 & 12.6 & 24.5 \\
\midrule
\textbf{Weighted avg.} & 2001 & 25.0 & 13.3 & 15.6 & \textbf{46.0} \\
\bottomrule
\end{tabular}
\end{table}

\begin{table}[!htbp]
\centering
\small
\setlength{\tabcolsep}{3pt}
\caption{LiveSWEAgent on SWE-bench Verified. Qwen3-8B and Qwen3-32B were excluded due to OpenRouter API errors with the bash-only LiveSWEAgent harness (Section~\ref{sec:experiments}, coverage note).}
\label{tab:failure_livesweagent_swebench}
\begin{tabular}{lrcccc}
\toprule
\textbf{Model} & \textbf{$n$} & \textbf{Search} & \textbf{Read} & \textbf{E.-Tgt.} & \textbf{E.-Qual.} \\
\midrule
GPT-5            & 176 &  5.1 & 11.9 & 15.3 & \textbf{67.6} \\
GPT-5-mini       & 214 &  8.4 & 13.1 & 15.4 & \textbf{63.1} \\
Qwen3-Coder-480B & 238 & 15.5 & \textbf{40.3} &  4.6 & 39.5 \\
Qwen3-Coder-30B  & 386 & 33.4 & 30.3 &  3.1 & 33.2 \\
Qwen3-235B       & 363 & 12.1 & 19.8 & 10.5 & \textbf{57.6} \\
\midrule
\textbf{Weighted avg.} & 1377 & 17.2 & 24.3 & 8.8 & \textbf{49.7} \\
\bottomrule
\end{tabular}
\end{table}

\begin{table}[!htbp]
\centering
\small
\setlength{\tabcolsep}{3pt}
\caption{SWE-Agent on PolyBench Verified (also reproduced inline as Table~\ref{tab:failure_sweagent_combined}, bottom).}
\label{tab:failure_sweagent_polybench}
\begin{tabular}{lrcccc}
\toprule
\textbf{Model} & \textbf{$n$} & \textbf{Search} & \textbf{Read} & \textbf{E.-Tgt.} & \textbf{E.-Qual.} \\
\midrule
GPT-5            & 208 & 18.8 & \textbf{41.3} &  8.7 & 31.2 \\
GPT-5-mini       & 216 & 27.8 & 30.1 &  8.8 & 33.3 \\
Qwen3-Coder-480B & 261 & 28.0 & \textbf{39.8} & 11.1 & 21.1 \\
Qwen3-Coder-30B  & 287 & 26.8 & \textbf{36.6} & 11.5 & 25.1 \\
Qwen3-235B       & 295 & \textbf{38.3} & 25.8 & 11.2 & 24.7 \\
Qwen3-32B        & 339 & \textbf{61.4} & 15.3 & 10.9 & 12.4 \\
Qwen3-8B         & 353 & \textbf{99.2} &  0.6 &  0.3 &  0.0 \\
\midrule
\textbf{Weighted avg.} & 1959 & \textbf{47.0} & 25.0 & 8.7 & 19.3 \\
\bottomrule
\end{tabular}
\end{table}

\begin{table}[!htbp]
\centering
\small
\setlength{\tabcolsep}{3pt}
\caption{OpenHands on PolyBench Verified.}
\label{tab:failure_openhands_polybench}
\begin{tabular}{lrcccc}
\toprule
\textbf{Model} & \textbf{$n$} & \textbf{Search} & \textbf{Read} & \textbf{E.-Tgt.} & \textbf{E.-Qual.} \\
\midrule
GPT-5            & 202 & 19.8 & \textbf{36.6} &  9.9 & 33.7 \\
GPT-5-mini       & 233 & 27.5 & 33.5 & 10.7 & 28.3 \\
Qwen3-Coder-480B & 244 & 24.2 & \textbf{39.8} &  5.7 & 30.3 \\
Qwen3-Coder-30B  & 293 & 25.6 & \textbf{38.9} &  6.1 & 29.4 \\
Qwen3-235B       & 301 & 30.9 & 33.6 &  7.6 & 27.9 \\
Qwen3-32B        & 348 & \textbf{52.9} & 25.9 &  2.9 & 18.4 \\
Qwen3-8B         & 361 & \textbf{65.7} & 18.6 &  4.4 & 11.4 \\
\midrule
\textbf{Weighted avg.} & 1982 & \textbf{37.9} & 31.3 & 6.4 & 24.4 \\
\bottomrule
\end{tabular}
\end{table}

\begin{table}[!htbp]
\centering
\small
\setlength{\tabcolsep}{3pt}
\caption{LiveSWEAgent on PolyBench Verified. Qwen3-8B and Qwen3-32B were excluded due to OpenRouter API errors with the bash-only LiveSWEAgent harness (Section~\ref{sec:experiments}, coverage note).}
\label{tab:failure_livesweagent_polybench}
\begin{tabular}{lrcccc}
\toprule
\textbf{Model} & \textbf{$n$} & \textbf{Search} & \textbf{Read} & \textbf{E.-Tgt.} & \textbf{E.-Qual.} \\
\midrule
GPT-5            & 239 & 16.7 & \textbf{54.8} &  4.6 & 23.8 \\
GPT-5-mini       & 279 & 20.8 & \textbf{53.8} &  6.1 & 19.4 \\
Qwen3-Coder-480B & 297 & 31.0 & \textbf{59.9} &  1.3 &  7.7 \\
Qwen3-Coder-30B  & 348 & \textbf{50.6} & 42.5 &  1.1 &  5.7 \\
Qwen3-235B       & 329 & 24.0 & \textbf{55.0} &  3.0 & 17.9 \\
\midrule
\textbf{Weighted avg.} & 1492 & 29.8 & \textbf{52.8} & 3.1 & 14.3 \\
\bottomrule
\end{tabular}
\end{table}

\paragraph{Cross-cell observations.} Three patterns are visible across the six tables. (1) On SWE-bench, capable models (GPT-5 family, Qwen3-Coder-480B/30B) consistently exhibit Edit-Quality as the modal failure mode (60--70\%), regardless of scaffold. (2) PolyBench shifts the modal failure category upstream: Search becomes the dominant mode for SWE-Agent (47.0\% weighted avg.), and Read for LiveSWEAgent (52.8\%), reflecting that multilingual repositories stress localization more than monolingual ones. (3) Smaller Qwen variants (8B, 32B) show consistently higher Search-failure rates than larger Qwen and GPT-5 models across every cell, supporting the interpretation that file localization is partially capability-bound.

\subsection{Dataset Statistics}

\begin{table}[!htbp]
\centering
\small
\caption{Complete dataset statistics.}
\label{tab:app_dataset_stats}
\begin{tabular}{lcc}
\toprule
\textbf{Metric} & \textbf{SWE-bench} & \textbf{PolyBench} \\
\midrule
Total instances & 500 & 382 \\
Languages & 1 (Python) & 4 (Py/Java/TS/JS) \\
Unique repositories & 12 & 20 \\
Avg files per patch & 1.4 & 1.6 \\
Avg functions per patch & 2.1 & 2.4 \\
Avg lines changed & 42 & 38 \\
\midrule
Models evaluated & 7 & 7 \\
Agents evaluated & 3 & 3 \\
Total trajectories & 9,500 & 7,258 \\
\bottomrule
\end{tabular}
\end{table}

\subsection{Assumptions and Limitations}
\label{sec:appendix_limitations}

\paragraph{Reference Patch Assumption.}
We assume the reference patch represents the canonical solution. Our convergence analysis (93.7\% exact function-level match, 100\% partial overlap) covers 350 instances solved by multiple models. A key question is whether harder, unsolved instances exhibit greater solution diversity. We cannot directly measure this since unsolved instances lack successful agent patches to compare. However, SWE-bench instances were designed to have well-defined solutions verified by test suites, suggesting canonical fix locations exist even for hard instances. Future work could incorporate alternative reference patches where available in the dataset, enabling multi-reference evaluation that computes recall against the union of valid fix locations.

\paragraph{Function Extraction.}
Function extraction uses tree-sitter parsing with regex fallback for robustness. While tree-sitter achieves high accuracy on well-formed code, the regex fallback may miss complex constructs such as nested functions, decorators, or metaprogramming patterns. The hybrid approach trades some accuracy for broad language support and handling of partial code snippets in trajectory observations.

\paragraph{Intervention Study Design.}
The two interventions in Section~\ref{sec:non_oracle} have specific scope: (1) \emph{Edit-commit checkpointing} is reported as a 5/5 existence proof on the EXACT-classified Near-Correct Corrupted cases (Appendix~\ref{sec:appendix_checkpoint_validation}); the realistic ceiling within Near-Correct Corrupted is bounded above by 30 (the EXACT + NEAR set) and remains an open empirical question, and a deployment-grade variant must gate the test invocation on cheaper signals to control cost. (2) \emph{Parallel-sample consensus} is measured end-to-end on a single (model, scaffold, benchmark) cell---GPT-5 + SWE-Agent + SWE-bench Verified, n=500---and the $+3.0$ pp result is sub-threshold at $\alpha = 0.05$ ($p = 0.08$); replication on Qwen3-Coder-480B and on the OpenHands and LiveSWEAgent scaffolds, and a held-out instance set to rule out leakage from the leave-one-out construction, are immediate follow-ups.

\paragraph{Benchmark Scope.}
Our evaluation focuses on SWE-bench and PolyBench, emphasizing bug fixes in well-maintained open-source repositories. Generalization to other task types (security vulnerabilities, performance optimization) or codebases (proprietary, poorly documented) requires further validation.

\subsection{Reproducibility and Artifact Release}
\label{sec:appendix_reproducibility}

We release the complete artifact at an anonymous repository link~\citep{anonymous_trajeval_artifact_2026}. The release is structured for direct reproduction of every figure and table in this paper:

\begin{itemize}
\item \textbf{Trajectory-extraction pipeline.} Per-architecture extractors for SWE-Agent (\texttt{view}/\texttt{str\_replace}), OpenHands (\texttt{fsRead}/\texttt{fsWrite}), and LiveSWEAgent (bash-only). Tree-sitter parsers for Python, Java, JavaScript, and TypeScript with a regex fallback.
\item \textbf{Per-trajectory feature CSV.} All 16,758 trajectories with stage-wise precision/recall, Pass@1, and assigned failure-mode bucket. A companion CSV (\texttt{failure\_mode\_breakdown\_tier1.csv}) provides the same fields recomputed against the Tier-1 call-graph-expanded reference for the sensitivity analysis in Appendix~\ref{sec:appendix_tier_sensitivity}, along with the recomputation script (\texttt{scripts/recompute\_tier1\_breakdown.py}). This is the single source of truth for every quantitative claim in the paper.
\item \textbf{Reproduction notebooks.} One per agent architecture, each running end-to-end from a small bundled trajectory sample to the corresponding row of Tables~\ref{tab:behavioral} and \ref{tab:failure_sweagent_combined}, so reviewers can verify the pipeline without re-executing 16,758 agent runs.
\item \textbf{Checkpoint validation artifacts.} The 5 EXACT intermediate-edit patches submitted to the SWE-bench Docker harness (\texttt{intermediate\_predictions\_dedup.jsonl}), per-instance test reports, and the structured 5/5 pass result (\texttt{final\_results.json}) for direct reproduction of Appendix~\ref{sec:appendix_checkpoint_validation}.
\end{itemize}

Random seed 42 is used throughout. The repository README documents which files reproduce which figures and tables in this paper.

\subsection{Edit-Quality Taxonomy: Assignment Rule and Threshold Robustness}
\label{sec:appendix_taxonomy_rule}

This appendix documents the cascading-precedence rule used to assign each Edit-Quality failure to exactly one sub-type (Section~\ref{sec:edit_quality_taxonomy}, Table~\ref{tab:edit_quality_taxonomy}), and reports a threshold-perturbation robustness check.

\paragraph{Assignment rule (MECE guarantee).} Each Edit-Quality failure is assigned to exactly one sub-type via a cascading precedence rule applied to three computed signals: (1) \emph{deletion overlap} ($\delta$): fraction of gold-patch deletion lines found in the agent's intermediate edits on the gold file; (2) \emph{token overlap} ($\tau$): fraction of gold-patch addition tokens reproduced in the agent's intermediate edits; (3) \emph{gold-file edit count} ($k$): number of distinct \texttt{str\_replace} operations the agent performed on gold file(s) during the trajectory.

The assignment proceeds as follows, with each case consuming the trajectory so that exactly one label is assigned:

\begin{enumerate}
    \item If $k = 0$ (agent never attempted an edit on the gold file): \textbf{Execution Gap}.
    \item If $\delta > 0.5$ and $\tau > 0.5$: \textbf{Near-Correct Corrupted} (the agent produced content highly similar to the gold patch, which was then either overwritten by a later edit or corrupted by a subsequent additive change).
    \item If $\delta > 0.5$ and $\tau \in (0.2, 0.5]$: classify as \textbf{Wrong API Call} (if gold and agent invoke disjoint method sets), \textbf{Missing Edge Case} (if gold adds branch logic absent from agent's edit), or \textbf{Right Location, Wrong Logic} (default).
    \item If $\delta \in (0.3, 0.5]$: \textbf{Incomplete Fix} (if gold is large and $k < 3$) or \textbf{Partial Understanding} (otherwise).
    \item If $k \geq 3$: \textbf{Confused Thrashing} (multiple attempts, none converging).
    \item Otherwise: \textbf{Wrong Branch Point} (if both gold and agent add control flow), \textbf{Wrong API Call} (if disjoint method sets), or \textbf{Wrong Function in File} (default).
\end{enumerate}

This cascading structure ensures mutual exclusivity: higher-overlap cases are consumed first (steps 2--4), then multi-attempt cases (step 5), with the remainder assigned by structural features (step 6). Thresholds ($\delta=0.5$, $\tau=0.5$, $k=3$) were selected via single-annotator inspection of a 50-trajectory calibration set. Inter-annotator reliability of the cascading rule's labels is reported separately in Appendix~\ref{sec:appendix_iaa}.

\paragraph{Threshold robustness.} To verify that the qualitative findings of Section~\ref{sec:edit_quality_taxonomy} are not artifacts of the chosen thresholds, we recomputed the theme distribution under $\pm 0.1$ perturbation of $\delta$ and $\tau$ and $\pm 1$ perturbation of $k$ (eight off-center configurations in addition to the default). Across all configurations:
\begin{itemize}
    \item Coherence Collapse remains the largest or second-largest theme, ranging 36--56\%.
    \item Scope/Completeness remains $< 10\%$ in every configuration.
    \item The relative ordering of model-level Coherence Collapse rates (Qwen3-32B largest, GPT-5/GPT-5-mini smallest) is preserved.
\end{itemize}
The Near-Correct Corrupted sub-type's exact-match validation (Appendix~\ref{sec:appendix_near_correct_validation}) is independent of these thresholds---it operates on diff-hunk reconstruction against the gold patch's structure---and is therefore unaffected.

\subsection{Inter-Annotator Agreement on the Cascading Rule}
\label{sec:appendix_iaa}

This appendix reports inter-annotator reliability of the Edit-Quality cascading rule's category labels (Section~\ref{sec:edit_quality_taxonomy}, Appendix~\ref{sec:appendix_taxonomy_rule}).

\paragraph{Procedure.} A second author independently labeled 100 SWE-bench Verified Edit-Quality failures, sampled uniformly at random from the 914-failure population. The annotator received only the agent trajectory (action history with observations) and the gold patch; they did not see the cascading rule's prior labels, the computed deletion/token/edit-count signals, or any other automated output. The annotator was given the category definitions from Appendix~\ref{sec:appendix_taxonomy_rule} (steps 1--6) and assigned each trajectory to exactly one category.

\paragraph{Pre-registration.} The cascading rule structure (steps 1--6 of Appendix~\ref{sec:appendix_taxonomy_rule}) and the threshold values ($\delta = 0.5$, $\tau = 0.5$, $k = 3$) were finalized before the second annotator began labeling; thresholds were not adjusted in response to the IAA outcome. The threshold-perturbation robustness check in Appendix~\ref{sec:appendix_taxonomy_rule} was likewise computed independently of the IAA labels. This separates the rule design from its reliability assessment: the second annotator's labels test the rule, they did not shape it.

\paragraph{Multi-level reporting.} We report agreement at three levels of granularity, chosen to match the paper's claim hierarchy rather than to maximize $\kappa$.

\begin{itemize}
    \item \textbf{3-class (headline).} Coherence Collapse vs.\ Semantic-or-Mislocalization vs.\ Execution Gap. This is the granularity at which the abstract and conclusion make claims (e.g., ``Coherence Collapse is the largest single theme''). \textbf{$\kappa = 0.80$, 90\% raw agreement.}
    \item \textbf{5-class (refinement).} 3-class scheme, plus separating Scope/Completeness from Semantic Error and preserving the two Coherence Collapse sub-types (Confused Thrashing vs.\ Near-Correct Corrupted). \textbf{$\kappa = 0.71$, 83\% raw agreement.}
    \item \textbf{6-category (full cascading rule).} The complete category set used in Table~\ref{tab:edit_quality_taxonomy}. \textbf{$\kappa = 0.68$, 77\% raw agreement.}
\end{itemize}

By the Landis--Koch convention, $\kappa = 0.80$ is the upper end of substantial (with ``almost perfect'' starting at $0.81$); $\kappa = 0.71$ is substantial; $\kappa = 0.68$ is at the lower end of substantial. The 3-class number is the one the paper's headline claim depends on.

\paragraph{Disagreement analysis.} Of the 23 fine-grained disagreements, 15 (65\%) involve the Scope/Completeness category---the rule's most contentious decision boundary, where step 4 of the cascading rule itself acts as a tie-break between Incomplete Fix and Partial Understanding. With direction tracked, the 15 Scope-related disagreements decompose as: 6 Rule=Scope/Human=Near-Correct Corrupted, 4 Rule=Scope/Human=Semantic Error, 3 Rule=Scope/Human=Confused Thrashing, and 2 Rule=Semantic/Human=Scope. The 6 Scope-to-NCC and 3 Scope-to-Thrashing disagreements are notable: in 9 of the 15 Scope-involving cases, the human annotator labels a trajectory as Coherence Collapse where the cascading rule labels it as Scope/Completeness, never the reverse. The 5 EXACT Near-Correct Corrupted cases on which the paper's checkpoint validation depends (Appendix~\ref{sec:appendix_near_correct_validation}) are validated by diff-hunk reconstruction and direct test-suite execution, both independent of the cascading rule's labels.

At the 3-class level, the 10 remaining disagreements all fall in the same direction: the cascading rule labels the trajectory in the Semantic-or-Mislocalization-or-Scope zone, while the second annotator reads it as Coherence Collapse. This directional asymmetry has a substantive interpretation: the cascading rule \emph{understates} rather than overstates Coherence Collapse. Reassigning all 10 disagreements to match the human annotator would only \emph{increase} the Coherence Collapse rate; the headline ``Coherence Collapse is the largest single theme'' is robust to either labeling choice, and arguably more so under the human's stricter classification.

\paragraph{Why we report all three levels.} A single $\kappa$ number masks where reliability is high vs.\ low. The full 9-sub-type taxonomy is retained in the paper's analysis for descriptive granularity, but the IAA we can defend most strongly is at the granularity of the central claim (3-class). Reporting all three levels allows readers to judge each claim against the appropriate reliability threshold.

\subsection{Near-Correct Corrupted: Diff-Hunk Validation}
\label{sec:appendix_near_correct_validation}

The 97 trajectories assigned to Near-Correct Corrupted by the rule above (Section~\ref{sec:edit_quality_taxonomy}) all exceed token-overlap thresholds $\delta > 0.5$ and $\tau > 0.5$. To strengthen this beyond bag-of-tokens overlap, we additionally ran a diff-hunk reconstruction check: for each case, we extracted the agent's intermediate edit at the moment of highest token overlap with the gold patch, normalized whitespace and stripped imports/comments, and compared its diff hunks against the gold patch's diff hunks. The result is reported in main-body Table~\ref{tab:near_correct_validation}: 30/97 (31\%) reconstruct gold-patch hunks at the strong-match level (EXACT or NEAR); 96/97 (99\%) target the gold patch's code region (everything except the single DIFFERENT case); the median gap between the near-correct edit and its overwriting/corrupting edit is 6 trajectory steps; and of the 5 EXACT cases, GPT-5 accounts for the largest share of strong-match cases per Edit-Quality failure (57\%), consistent with the broader pattern that capable models reach correct fixes more often before losing them.

\subsection{Edit-Commit Checkpoint Validation}
\label{sec:appendix_checkpoint_validation}

This appendix reports the validation underlying Prediction 1 (Section~\ref{sec:non_oracle}). The 5 EXACT-classified intermediate edits identified by the diff-hunk validation (Appendix~\ref{sec:appendix_near_correct_validation}) are by construction bit-identical (modulo whitespace) to the gold patch's hunks. We tested whether they actually pass the existing test suite, which would establish edit-commit checkpointing as a deployable intervention rather than a hypothetical one.

\paragraph{Setup.} For each of the 5 EXACT cases, we extracted the agent's intermediate edit at the trajectory turn of highest gold-token overlap, reconstructed the resulting candidate patch, and submitted it to the official SWE-bench Verified Docker harness. The harness applies the patch to the pre-bug commit and runs the full instance test suite; a case is considered recovered if and only if all FAIL\_TO\_PASS tests pass and no PASS\_TO\_PASS tests regress.

\paragraph{Result.} All 5 EXACT cases pass the full SWE-bench test suite. This confirms that these agents produced a provably correct solution mid-trajectory that was subsequently corrupted or overwritten by a later edit. An edit-commit checkpoint that freezes test-passing intermediate edits would have recovered all 5 instances. We report this as a strict existence proof---the 5 EXACT cases are the subset where the intermediate edit is bit-identical to the gold patch and demonstrably passes the full test suite, establishing edit-commit checkpointing as a deployable intervention rather than a hypothetical one. The realistic ceiling within Near-Correct Corrupted is bounded above by 30 (the EXACT + NEAR set, 31\% of the 97 NCC cases) and remains an open empirical question.

\paragraph{Cost model.} Edit-commit checkpointing is not free. A naive implementation runs the FAIL\_TO\_PASS test subset after every \texttt{str\_replace} on a gold-area file, which can multiply per-trajectory test invocations by 10--40$\times$ depending on edit count and test-suite duration. Practical deployments would gate the test invocation on cheaper signals (e.g., diff-similarity to prior edits, or a learned predictor over edit content) to avoid running the harness on edits unlikely to be improvements. We report the 5/5 result under the strictest possible cost assumption (test every candidate edit) so that the existence-proof claim is not contingent on cost-saving heuristics; deployment-grade variants are an open engineering question.

\paragraph{Scope and limitations.} The validation establishes a strict lower bound: the 5 EXACT cases are the subset where the intermediate edit is bit-identical to the gold patch and passes tests \emph{by construction}. We attempted broader validation on the 25 NEAR cases but our current diff-reconstruction pipeline introduces patch-application errors (missing line-number context) that confound semantic-correctness measurement; we leave end-to-end agent re-execution under checkpointing to follow-up work, where the agent would be re-run with a real test-success-triggered freeze rather than retroactively evaluated. The 5 EXACT cases are sufficient to establish that edit-commit checkpointing is a deployable, not merely hypothetical, intervention; the realistic ceiling within Near-Correct Corrupted is bounded above by 30 (the EXACT + NEAR set) and remains an open empirical question.

\subsection{Consensus-as-Relevance-Signal Validation}
\label{sec:appendix_consensus}

This appendix reports the validation underlying Prediction 2 (Section~\ref{sec:non_oracle}). The oracle-guided intervention in Section~\ref{sec:results} requires the reference patch at execution time and is therefore not deployable. We test whether a non-oracle relevance signal, computed from parallel-sample agreement among models alone, recovers a comparable lift.

\paragraph{Signal definition.} For each SWE-bench Verified instance $i$ and threshold $N \in \{2, 3, 4, 5\}$, we define
$$\textsc{Consensus}_N(i) = \{f \mid f \text{ viewed by} \geq N \text{ of the 6 non-GPT-5 models}\},$$
i.e., the leave-one-out set of files that at least $N$ of the other six models inspected during their trajectories on instance $i$. The signal uses no reference patch and no information from GPT-5 itself.

\paragraph{Recovery test.} Of the 500 SWE-bench Verified instances, the oracle-guided intervention (Section~\ref{sec:results}) would fire on 490 (the cases where GPT-5 viewed at least one gold file). For each of these 490 instances we ask: does $\textsc{Consensus}_N$ contain at least one gold file? If yes, the consensus signal would have triggered the same confirmatory feedback that produced the +4.6 pp oracle gain.

\paragraph{Results.} Table~\ref{tab:consensus_validation} reports precision and recall of $\textsc{Consensus}_N$ against gold files (averaged across 500 instances), the fraction of the 490 oracle-firing cases that consensus also covers, and the implied lower bound on Pass@1 lift (recovery $\times$ 4.6 pp).

\begin{table}[!htbp]
\centering
\small
\caption{Consensus signal validation. For each threshold $N$, we compute the set of files viewed by $\geq N$ of the 6 non-GPT-5 models on each instance. Precision/recall are against gold edit locations. ``Oracle recovered'' is the fraction of the 490 instances where the oracle would fire that consensus also covers at least one gold file. Analytical ceiling = recovery rate $\times$ 4.6 pp (the oracle gain), achieved only under perfect-precision triggering.}
\label{tab:consensus_validation}
\begin{tabular}{ccccc}
\toprule
$N$ & Precision & Recall & Oracle recovered & Analytical ceiling \\
\midrule
2 & 35.2\% & 95.5\% & 99.0\% (485/490) & +4.6 pp \\
3 & 52.4\% & 94.2\% & 97.3\% (477/490) & +4.5 pp \\
4 & 66.8\% & 91.5\% & 92.4\% (453/490) & +4.3 pp \\
5 & 75.6\% & 86.2\% & 82.4\% (398/483) & +3.8 pp \\
\bottomrule
\end{tabular}
\end{table}

\paragraph{Analytical interpretation.} At $N{=}3$, $\textsc{Consensus}_N$ recovers 97.3\% of oracle-triggering cases. Multiplying by the $+4.6$ pp oracle ceiling implies an analytical ceiling of $+4.5$ pp Pass@1 lift, achieved only if every consensus-driven trigger produced the same gain as an oracle-driven trigger (which requires perfect-precision triggering---a condition the 52.4\% precision at $N{=}3$ does not meet). Increasing $N$ trades recovery for precision: at $N{=}5$, precision rises to 75.6\% but recovery drops to 82.4\%, suggesting $N{=}3$ is a reasonable operating point.

\paragraph{End-to-end measurement.} The end-to-end consensus measurement is reported in main-body Table~\ref{tab:consensus_endtoend} (Section~\ref{sec:non_oracle}). At $\textsc{Consensus}_3$ precision of 52.4\%, the gain/loss decomposition (39 gained, 24 lost, net +15) reflects that roughly half of the confirmatory triggers occur on files outside the gold set; the agent occasionally over-weights these confirmations and edits the wrong location. This identifies a concrete mechanism for the gap to the analytical bound and suggests a direct improvement path: combining $\textsc{Consensus}_N$ with a higher-precision filter (e.g., $N{=}4$ or a learned re-ranker over consensus-flagged files) to suppress losses without sacrificing gains.

\paragraph{Implications.} The result establishes that the Edit-Quality bottleneck identified by \codename{} is not gated on having a reference patch at deployment. The function-level convergence we observe (93.7\% on multiply-resolved instances, Section~\ref{sec:results}) is empirically strong enough that consensus among independent samples reproduces the oracle's confirmatory signal almost entirely. Self-consistency decoding~\citep{wang2022self} is the chain-of-thought analogue; here it operates on tool-use trajectories rather than token sequences. Two caveats apply: (i) consensus requires running $\geq N{+}1$ samples, which has compute cost (this matches self-consistency); (ii) the validation is conducted on the same 500 SWE-bench Verified instances on which the per-model trajectories were collected, so we cannot rule out leakage from instance-level features into the consensus signal---though the leave-one-out construction (excluding GPT-5 itself) and the fact that no reference patch enters the signal partially mitigate this. Re-running consensus on a held-out instance set, and tightening the gap between the empirical $+3.0$ pp measurement and the analytical $+4.5$ pp ceiling via higher-precision triggering, are immediate follow-ups.

\subsection{Trajectory Length and Coherence Collapse}
\label{sec:appendix_traj_length}

A natural concern is whether Coherence Collapse is simply long-context degradation under a different name---agents that run longer exhaust their context window and produce incoherent output. We test this by computing Spearman rank correlations between trajectory length (number of history steps) and failure sub-type assignment across all 914 Edit-Quality failures (SWE-Agent, SWE-bench Verified).

\begin{figure*}[t]
\centering
\includegraphics[width=\linewidth]{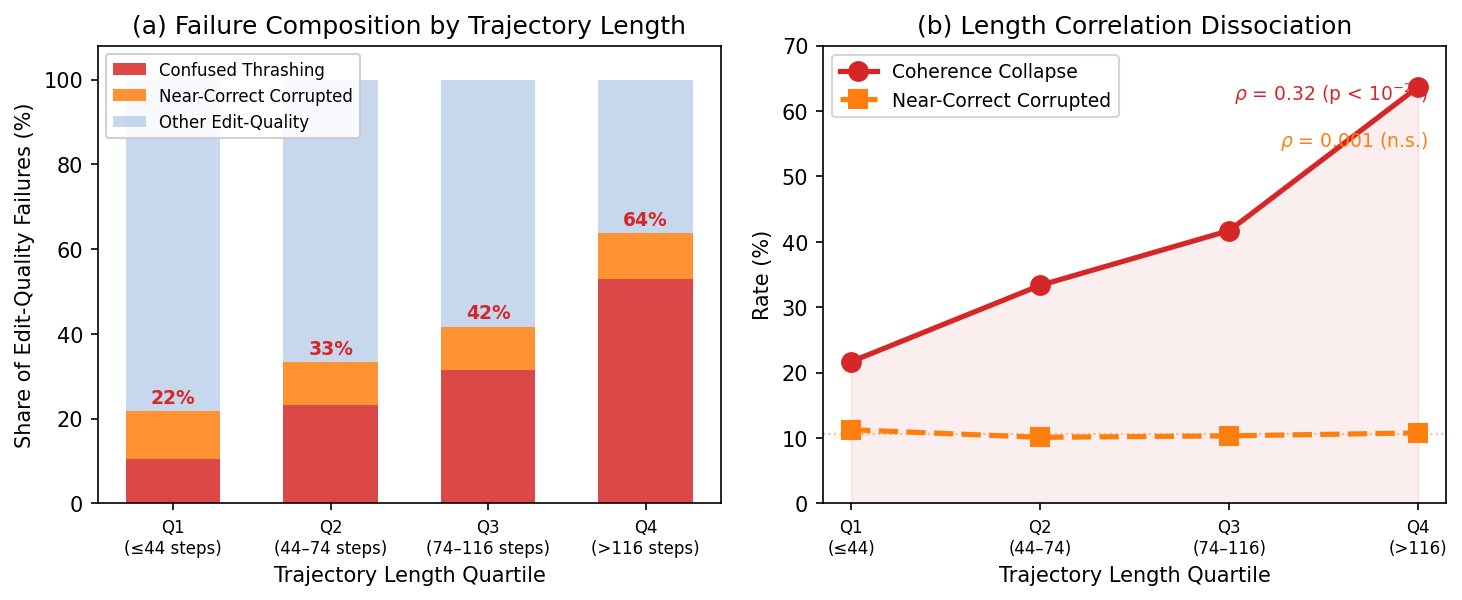}
\caption{Relationship between trajectory length and Coherence Collapse. (a) Failure theme composition by trajectory-length quartile; Coherence Collapse (red+orange) grows from 22\% to 64\% (aggregate $\rho = 0.32$). (b) Dissociation: Confused Thrashing (red) scales with length ($\rho = 0.34$) while Near-Correct Corrupted (orange) is flat ($\rho = 0.001$).}
\label{fig:traj_length}
\end{figure*}

\paragraph{Results (Figure~\ref{fig:traj_length}).}
Coherence Collapse shows a moderate positive correlation with trajectory length ($\rho = 0.32$, $p = 3.4 \times 10^{-23}$): the Q4 (longest) quartile has a 63.7\% collapse rate versus 21.7\% in Q1 (shortest), a 42 pp difference. However, the sub-type decomposition reveals a critical dissociation: \emph{Near-Correct Corrupted} shows \textbf{no correlation} with trajectory length ($\rho = 0.001$, $p = 0.98$), maintaining a stable $\sim$10--11\% rate across all quartiles. The correlation is driven almost entirely by \emph{Confused Thrashing} (the multi-attempt variant; $\rho = 0.34$, $p = 8.5 \times 10^{-27}$), which rises from 10.5\% in Q1 to 52.9\% in Q4.

Among the 97 Near-Correct Corrupted cases, the gap between the near-correct intermediate edit and its corruption is strongly correlated with total trajectory length ($\rho = 0.77$, $p < 10^{-14}$, $n = 64$), indicating that longer trajectories introduce more intervening steps before corruption---but do not increase the \emph{probability} of corruption.

\paragraph{Interpretation.} Coherence Collapse is partially but not entirely explained by long-context effects. The Confused Thrashing variant (29.1\% of Edit-Quality failures) is length-dependent---these agents likely run longer \emph{because} they are thrashing, not the reverse (the causal direction is confounded by the data). The Near-Correct Corrupted variant (10.6\%) is length-independent, suggesting a qualitatively different failure mechanism: the agent produces a correct fix early in the trajectory regardless of total length, then corrupts it through overconfident refinement. This dissociation indicates that context-window degradation is not the sole driver of Coherence Collapse and supports the taxonomy's distinction between the two sub-types.

\subsection{Edit-Quality Worked Examples}
\label{sec:appendix_worked_examples}

We present three trajectories illustrating the dominant Coherence Collapse sub-types, drawn from our analysis of the 914 Edit-Quality failures.

\subsubsection{Worked Example 1: Confused Thrashing}
\label{sec:example_thrashing}

\textbf{Instance:} \texttt{django\_\_django-11848} \quad \textbf{Model:} GPT-5 \quad \textbf{Gold file:} \texttt{django/utils/http.py}

\textbf{Gold patch} (two-year window fix for HTTP date parsing):
\begin{lstlisting}[basicstyle=\ttfamily\tiny]
-            if year < 70:
-                year += 2000
+            current_year = datetime.datetime.utcnow().year
+            current_century = current_year - (current_year % 100)
+            if year - (current_year % 100) > 50:
+                year += current_century - 100
             else:
-                year += 1900
+                year += current_century
\end{lstlisting}

\textbf{Agent trajectory} (4 edits on gold file, none preserved):
\begin{lstlisting}[basicstyle=\ttfamily\tiny]
Turn 22: str_replace http.py
  OLD: "if year < 70: year += 2000 else: year += 1900"
  NEW: (replaces with month/day extraction logic -- wrong approach)

Turn 28: str_replace http.py
  OLD: (reverts and refactors year parsing)
  NEW: "year_str = m.group('year'); year = int(year_str)..."
       (adds variable but loses the century-window logic)

Turn 34: str_replace http.py  [FAILED -- old_str not found]

Turn 42: str_replace http.py  [FAILED -- old_str not found]
\end{lstlisting}

\textbf{Final patch:} Edits \texttt{django/db/migrations/serializer.py} (completely unrelated file).

\textbf{Diagnosis:} The agent found the correct file and twice attempted the correct code region. However, each edit overwrote the previous without converging on the gold logic. After two failed \texttt{str\_replace} attempts (stale \texttt{old\_str} due to prior edits), the agent abandoned the approach entirely and pivoted to an unrelated fix. This is the signature \emph{Confused Thrashing} pattern: the correct location is reached but coherent multi-step reasoning collapses.

\subsubsection{Worked Example 2: Near-Correct Corrupted (Additive)}
\label{sec:example_near_correct}

\textbf{Instance:} \texttt{matplotlib\_\_matplotlib-20859} \quad \textbf{Model:} GPT-5 \quad \textbf{Gold file:} \texttt{lib/matplotlib/legend.py}

\textbf{Gold patch} (support legends on SubFigure by importing FigureBase):
\begin{lstlisting}[basicstyle=\ttfamily\tiny]
-        from matplotlib.figure import Figure
+        from matplotlib.figure import FigureBase
         ...
-        elif isinstance(parent, Figure):
+        elif isinstance(parent, FigureBase):
\end{lstlisting}

\textbf{Agent trajectory:}
\begin{lstlisting}[basicstyle=\ttfamily\tiny]
Turn 48: str_replace legend.py            <-- CORRECT
  OLD: "from matplotlib.figure import Figure"
  NEW: "from matplotlib.figure import FigureBase"

Turn 70: str_replace legend.py            <-- CORRUPTS (additive)
  OLD: "elif isinstance(parent, Figure): ..."
  NEW: "elif isinstance(parent, FigureBase):
            self.set_figure(parent.figure)  ..."
       (adds .figure indirection -- not in gold)
\end{lstlisting}

\textbf{Final patch:} Contains \emph{both} edits, but the turn-70 change adds \texttt{parent.figure} instead of the gold's direct reference, causing the test suite to fail.

\textbf{Diagnosis:} At turn 48, the agent produces an edit that \emph{exactly matches} the gold patch's first hunk. At turn 70, a second edit is semantically close but adds an extra \texttt{.figure} dereference not present in the gold. The import change alone would have sufficed; the over-correction at turn 70 introduced the failure. This is the \emph{additive corruption} variant of Near-Correct Corrupted: the original correct edit is preserved, but a subsequent well-intentioned refinement breaks the patch.

\subsubsection{Worked Example 3: Near-Correct Corrupted (Destructive Overwrite)}
\label{sec:example_exact_overwrite}

\textbf{Instance:} \texttt{django\_\_django-15277} \quad \textbf{Model:} Qwen3-32B \quad \textbf{Gold file:} \texttt{django/db/models/fields/\_\_init\_\_.py}

\textbf{Gold patch} (guard \texttt{MaxLengthValidator} against \texttt{None}):
\begin{lstlisting}[basicstyle=\ttfamily\small]
-        self.validators.append(validators.MaxLengthValidator(self.max_length))
+        if self.max_length is not None:
+            self.validators.append(validators.MaxLengthValidator(self.max_length))
\end{lstlisting}

\textbf{Agent trajectory:}
\begin{lstlisting}[basicstyle=\ttfamily\small]
Turn 34: str_replace fields/__init__.py       <-- EXACT GOLD EDIT
  OLD: "self.validators.append(validators.MaxLengthValidator(self.max_length))"
  NEW: "if self.max_length is not None:
            self.validators.append(validators.MaxLengthValidator(self.max_length))"

Turn 48: str_replace fields/__init__.py       <-- DESTRUCTIVE OVERWRITE
  OLD: "self.db_collation = db_collation
        self.validators.append(validators.MaxLengthValidator(self.max_length))"
  NEW: "if self.max_length is not None:
            self.validators.append(validators.MaxLengthValidator(self.max_length))"
       (deletes `self.db_collation = db_collation` -- breaks CharField)
\end{lstlisting}

\textbf{Final patch:} Only contains a \texttt{reproduce\_error.py} script; the source-file edits cancelled out or broke the build, yielding no productive diff.

\textbf{Diagnosis:} At turn 34, the agent produces a patch \emph{identical} to the gold reference (\texttt{del\_recall}=1.00, \texttt{add\_recall}=1.00, \texttt{add\_seq}=1.00). We confirmed via the SWE-bench Docker harness that the turn-34 patch passes the full test suite (Appendix~\ref{sec:appendix_checkpoint_validation}). Fourteen steps later, the agent re-attempts the same logical fix but with a larger \texttt{old\_str} context that inadvertently deletes the \texttt{db\_collation} assignment. This destroys the correct turn-34 edit and breaks \texttt{CharField} initialization. The agent had already solved the problem---the failure was caused purely by an unnecessary follow-up edit. This is the \emph{destructive overwrite} variant of Near-Correct Corrupted: the test-passing intermediate edit is deleted and replaced by a corrupted version.

\end{document}